\documentclass[aps, prb, twocolumn, superscriptaddress, english]{revtex4-2}

\usepackage{graphicx}
\usepackage{cancel}
\usepackage{enumerate}
\usepackage{hyperref}
\usepackage{textcomp}
\usepackage{amsmath}
\usepackage{amssymb}
\usepackage{pgf}
\usepackage{soul}
\usepackage{subfigure}
\usepackage[english]{babel}
\usepackage[normalem]{ulem}

\def\ket#1{| #1\rangle}

\def\ee{\mathrm{e}}

\newcommand{\Ec}{E_c}

\newcommand{\J}{{J_0}}

\begin{document}

\title{Engineering a Josephson junction chain for the simulation of the quantum clock model}
\author{Matteo M. Wauters}
  \affiliation{CNR-INO Pitaevskii BEC Center and Dipartimento di Fisica, Universit\`a di Trento,  Via Sommarive 14, I-38123 Trento, Italy}
 \affiliation{INFN-TIFPA, Trento Institute for Fundamental Physics and Applications, Via Sommarive 14, I-38123 Povo, Trento, Italy}
\affiliation{Niels Bohr International Academy and Center for Quantum Devices, Niels Bohr Institute, Copenhagen University, Universitetsparken 5, 2100 Copenhagen, Denmark}

\author{Lorenzo Maffi}
\affiliation{Dipartimento di Fisica e Astronomia ``G. Galilei'', Via Marzolo 8, I-35131 Padova, Italy}
\affiliation{INFN, Sezione Padova, I-35131 Padova, Italy}
\affiliation{Niels Bohr International Academy and Center for Quantum Devices, Niels Bohr Institute, Copenhagen University, Universitetsparken 5, 2100 Copenhagen, Denmark}

\author{Michele Burrello}
\affiliation{Niels Bohr International Academy and Center for Quantum Devices, Niels Bohr Institute, Copenhagen University, Universitetsparken 5, 2100 Copenhagen, Denmark}
\affiliation{Dipartimento di Fisica dell’Università di Pisa, Largo Pontecorvo 3, I-56127 Pisa, Italy}

\begin{abstract}
The continuous improvement of fabrication techniques and high-quality semiconductor-superconductor interfaces allows for an unprecedented tunability of Josephson junction arrays (JJA), making them a promising candidate for analog quantum simulations of many-body phenomena. While most experimental proposals so far focused on quantum simulations of ensembles of two-level systems, the possibility of tuning the current-phase relation beyond the sinusoidal regime paves the way for studying statistical physics models with larger local Hilbert spaces. Here, we investigate a particular JJA architecture that can be mapped into a $\mathbb{Z}_3$ clock model. Through matrix-product-states simulations and bosonization analysis, we show that few experimentally accessible control parameters allow for the exploration of the rich phase diagrams of the associated low-energy field theories. 
Our results expand the horizon for analog quantum simulations with JJAs towards models that can not be efficiently captured with qubit architectures.
\end{abstract}

\maketitle

\section{Introduction}\label{sec:intro}

The steady development of quantum technologies opened the path to engineer systems for the realization of many-body quantum phenomena that, on one side, are considered crucial for the progress of quantum many-body theory but, on the other, are so elusive to be hardly observable in natural systems. In particular, the advances of analog quantum simulators in trapped ultracold gases, ions, and Rydberg atoms sparked a renewed interest in exploring models based on discrete degrees of freedom and symmetries. These include, for instance, lattice gauge theories with finite gauge groups \cite{zohar2015,banuls2020,banuls2020b} or quantum spin liquids \cite{semeghini2021}.

In this framework, models with $\mathbb{Z}_N$ symmetries play an important role because they offer the possibility to interpolate between the physics of $\mathbb{Z}_2$-symmetric models, typically easier to study both analytically and numerically, and problems with continuous symmetries. $\mathbb{Z}_3$-symmetric models, in particular, attracted considerable attention and offer one of the most common playgrounds to study chiral phases in quantum one-dimensional systems \cite{Zhuang2015,Whitsitt2018,Samajdar2018}. States with a spontaneous breaking of a $\mathbb{Z}_3$ symmetry emerge in Rydberg atom chains \cite{bernien2017,Keesling2019} where they determine the appearance of non-conformal and chiral quantum phase transitions \cite{Fendley2004,Chepiga2019,Giudici2019,Chepiga2021}. Additionally, $\mathbb{Z}_3$ quantum clock models \cite{Ostlund1981,Howes1983} have been used to describe the physics of parafermionic zero-energy modes \cite{fendley2012,Milsted2014,cobanera2016,iemini2017,Moran2017,mazza2018,rossini2019,Mahyaeh2020,Wouters2022} and to design systems with exotic topological order and anyons \cite{Alicea2014,Mong2014,Alicea2015}.

Here we propose a solid-state realization of quantum clock models in one dimension based on hybrid semiconductor-superconductor devices \cite{krogstrup2015,shabani2016}. These platforms combine the coherence of superconducting (SC) systems with the tunability via electrostatic gates offered by their semiconducting substrate \cite{shabani2016,Kjaergaard2017,Casparis_NatNanoTech2018,Ciaccia2023}. Moreover, their fabrication techniques allow for considerable scalability, with systems of thousands of Josephson junctions studied in recent experiments \cite{bottcher2018,manucharyan2019,higginbotham2022}.

These experimental architectures offer an ideal platform to explore in a controllable way fundamental quantum many-body phenomena, as BKT transitions \cite{bottcher2018,bottcher2022,Bottcher2022b}, and to investigate the properties of emergent quantum field theories \cite{Roy2019,Roy2023.1}, including integrable \cite{Saleur2021} and conformal models~\cite{Roy2023.2,maffi2023}.

In this work, we take advantage of their tunability to design SC chains with an emergent $\mathbb{Z}_3$ symmetry. The platform that we design presents a very rich phase diagram that enables the observation of non-trivial quantum critical states including conformal Potts phase transitions, tricritical Lifshitz lines, and gapless floating phases \cite{Ostlund1981,Howes1983}.
A few experimentally accessible global parameters span this phase diagram, which we characterize through a combination of field-theory approaches based on bosonization and matrix-product-states (MPS) numerical techniques optimized with the density matrix renormalization group (DMRG) algorithm. 

Our construction is based on a building block that relies on the non-sinusoidal energy-phase relation typical of hybrid Josephson junctions. In Sec. \ref{sec:TJJ} we describe a triple Josephson junction that can be operated as a quantum clock degree of freedom when tuned in a transmon-like regime.
Based on these elementary units, we derive the Hamiltonian for the Josephson junction array (JJA) in Sec.~\ref{sec:JJA} and its mapping on $\mathbb{Z}_3$ clock models, while in Sec.~\ref{sec:ft} we discuss its low-energy effective field-theory description.
Section \ref{sec:results} contains the bulk of our numerical results. We first analyze the Potts phase transition in Sec.~\ref{sec:potts} and its classical limit in Sec.~\ref{sec:class}, then we characterize the phase diagram of the chiral clock model in Sec.~\ref{sec:chiral}.
In Sec.~\ref{sec:ng} we describe the role played by the induced charge on the SC islands and we discuss the deviations from the $\mathbb{Z}_3$-symmetric case and the effect of disorder in Sec.~\ref{sec:pert_dmrg}.
Finally, we draw our conclusions and discuss possible outlooks in Sec.~\ref{sec:conclusion}.
Further analytical and numerical analysis are presented in the Appendix.

\section{A triple Josephson junction operated as a quantum clock} \label{sec:TJJ}

\begin{figure}
    \centering
    \includegraphics[width=\linewidth]{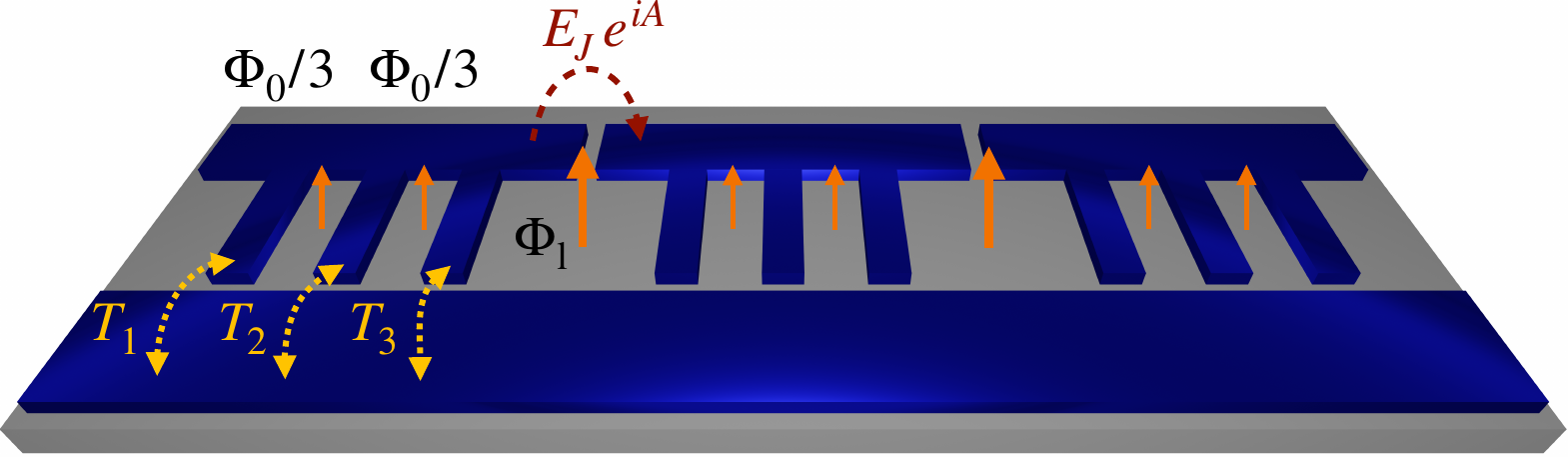}
    \caption{Sketch of the JJA used to simulate the $\mathbb{Z}_3$ clock Hamiltonian. Each SC island (blue) is coupled to a common grounded SC background through three parallel tunable Josephson junctions with transparencies $T_p$. Cooper pairs can additionally tunnel between neighboring islands via sinusoidal Josephson junctions with energy $E_J$.}
    \label{fig_device} 
\end{figure}

The first task to design a one-dimensional Josephson junction chain for the quantum simulation of models with a global $\mathbb{Z}_3$ symmetry is the definition of a suitable elementary block that can be operated as a 3-state quantum clock (or, equivalently, as a qutrit). Therefore we must consider an elementary SC element that, in the ideal limit, displays three degenerate ground states, which we label by $\ket{s}$ with $s=0,1,2$. It is useful to introduce a pair of clock operators, $\sigma$ and $\tau$, acting on these three states in the following way:
\begin{equation}
\sigma \ket{s} = \ee^{i\frac{2\pi s}{3}} \ket{s} \,,\quad \tau \ket{s} = \ket{(s-1) \mod 3}\,.
\end{equation}
The operators $\sigma$ and $\sigma^\dag$ distinguish the three ground states, whereas the operators $\tau$ and $\tau^\dag$ represent equal-amplitude transition among them, such that:
\begin{equation} \label{comm}
\tau \sigma = \ee^{-i \frac{2\pi}{3}}\sigma \tau\,.
\end{equation}

In order to design $\mathbb{Z}_3$ symmetric models we must impose that the dynamics of a single SC building block can be expressed purely as a function of the operators $\tau$ and $\tau^\dag$, such that, indeed, the three states $\ket{s}$ are equivalent.

To this purpose, we consider a floating SC island connected with a background superconductor via a triple Josephson junction, see Fig.~\ref{fig_device}, that displays the following Josephson potential:
\begin{equation} \label{pot}
V(\varphi) = -\J \cos 3 \varphi \,,
\end{equation}
where $\varphi$ is the operator that represents the phase difference between the SC island and the background superconductor, such that $\ee^{i\varphi} \sim \sigma$ describes the tunneling of a single Cooper pair from the floating SC island to the background superconductor. The potential in Eq. \eqref{pot} models, therefore, the coherent tunneling of triplets of Cooper pairs to and from the floating island. For its implementation, it is necessary to suppress the tunneling of single Cooper pairs, and of pairs of Cooper pairs, which would correspond to potentials terms of the kind $\cos{\varphi}$ and $\cos{2\varphi}$ respectively. The suppression of these lower harmonics can be obtained by destructive interference when considering three non-sinusoidal Josephson junctions in parallel, as presented in Ref. \cite{maffi2023}. In particular, we consider Josephson junctions constructed with metallic islands epitaxially grown over a semiconducting material \cite{shabani2016}. In these setups, the tunneling of Cooper pairs across the junction is mediated by Andreev states localized on the semiconducting region surrounding the junction. A single Josephson junction of this kind, which we label by $p$, displays an energy-phase relation of the kind \cite{beenakker1991}:
\begin{equation}
\label{beenakker}
\mathcal{E}_J^{(p)}\left(\varphi\right)=-\Delta\sqrt{1-T_p\sin^2{\left(\varphi/2\right)}}\,,
\end{equation}
where we are assuming that the Cooper pair tunneling is mediated by a single Andreev state; $0<T_p<1$ represents the transparency of the junction and $\Delta$ is the SC gap induced in the semiconducting 2DEG. 

To engineer the ideal potential \eqref{pot}, we consider a device with three identical Josephson junctions, such that $T_1=T_2=T_3$, arranged to form two SC loops with the same area (Fig.~\ref{fig_device}).

Through an out-of-plane magnetic field, we introduce the same magnetic flux $\Phi$ in both loops, such that $\Phi/\Phi_0 = 1/3$, where $\Phi_0=hc/2e$ is the magnetic flux quantum. Under these conditions, analogously to Ref. \cite{maffi2023}, we obtain that the total potential $\mathcal{E}_J^{(1)}\left(\varphi-\frac{2\pi}{3}\right) + \mathcal{E}_J^{(2)}\left(\varphi\right) +\mathcal{E}_J^{(3)}\left(\varphi+\frac{2\pi}{3}\right)$ can be approximated by the form in Eq. \eqref{pot}.  Perturbations around this optimal point will result in the splitting of the minima of the potential, which we will address in Sec. \ref{sec:pert_dmrg}.

Concerning the experimental realization of this symmetric triple junction, we emphasize that the general scenario with Josephson junctions characterized by multiple Andreev states does not yield qualitative differences to implement the potential \eqref{pot}: the addition of fluxes $\Phi_0/3$ makes indeed the potential periodic in $\varphi$ with periodicity $2\pi/3$, with a dominant $\cos 3\varphi$ component. In this respect, more general descriptions of the energy-phase relation \eqref{beenakker} can be adopted \cite{beenakker1991,kruti2024} without affecting the required symmetries.

Josephson junctions with the dispersion $\mathcal{E}_J^{(p)}$ can also be realized with alternative techniques. For instance, recent studies have shown that the energy-phase relation \eqref{beenakker} can be efficiently achieved by considering two traditional junctions in series \cite{Bozkurt2023,Banszerus2024}. A key aspect is that in hybrid semiconductor-superconductor platforms, their transparency $T_p$ can be tuned via suitable electrostatic gates \cite{Kjaergaard2017,Casparis_NatNanoTech2018,Ciaccia2023}, 
and, in turn, it determines the ratio between the harmonics $\cos n \varphi$ appearing in $\mathcal{E}_J$.

The resulting construction of 3-state quantum clocks based on three parallel SC arms, each composed of two sinusoidal Josephson junctions, is reminiscent of the so-called masus devices \cite{Doucot2004}. The controllability of the external transparencies $T_1$ and $T_3$ makes it possible to implement the potential \eqref{pot}, and, very recently, their two-arm counterparts, commonly called Josephson rhombi \cite{Gladchenko2008}, have been successfully realized in tunable hybrid InAs - Al devices \cite{banszerus2024b}.

So far, we considered only the Josephson potential of the triple junction. We now introduce the effect of its charging energy of this device by considering the following Hamiltonian:
\begin{equation} \label{ham2}
H_{\rm TJJ}=\Ec\left(N - n_g\right)^2 + \J\left(1-\cos 3 \varphi\right)\,.
\end{equation}
Here $N$ is the operator that represents the number of excess Cooper pairs in the floating island, such that $\left[N,\ee^{i\varphi}\right]=\ee^{i\varphi}$ and we can consider $N=-i\partial_\varphi$. $n_g$ is the charge induced by external potentials on the SC island, in units of $2e$; finally, the charging energy of this element is given by $\Ec=2e^2/C_{\rm tot}$, where $C_{\rm tot}$ is the total self-capacitance of the island. 

To understand the physics of the triple Josephson junction we can consider different limits. When $\J$ dominates over $\Ec$, this system lies in a transmon-like regime: the three degenerate minima of the potential $V(\varphi)$ determine the three ground states $\ket{s}$ whose wavefunctions are peaked around $\varphi=0,2\pi/3$ and $4\pi/3$. Hence, the triple junction represents a quantum clock degree of freedom labeled by the operator $\sigma \sim \ee^{i\varphi}$. The charging energy plays the role of a kinetic term which causes phase slips between these three ground states. Analogously to transmons \cite{Koch2007}, the related amplitudes can be estimated based on a semiclassical approximation and the effective low-energy Hamiltonian corresponds to the following description in terms of the $\tau$ clock operator:
\begin{multline} \label{slip}
H_{\rm eff}\overset{\J > \Ec }{\approx}-\alpha \Ec^{1/4}\J^{3/4} \ee^{-\sqrt{\frac{32 \J}{9\Ec}}+ in_g \frac{2\pi}{3}} \tau + {\rm H.c.} \\ 
\equiv -h_\tau \left(\ee^{i n_g \frac{2\pi}{3}} \tau + \ee^{-i n_g \frac {2\pi}{3}} \tau^\dag\right)\,,
\end{multline} 
with $\alpha\approx 18.6$ a numerical constant (see Appendix \ref{app:WKB}). In the regime $\J \gg \Ec$, the amplitude of the $2\pi/3$ phase slips is exponentially suppressed in $\J/\Ec$ and the three states are separated from higher energy states by a gap proportional to the plasma frequency $\sqrt{\J \Ec}$. 

For increasing $\Ec$, however, the splitting of the three lowest-energy states caused by the phase slips becomes of order $\sim \Ec$ and the three-level approximation fails if $\Ec \sim \J$. By increasing $\Ec$ further, the triple junction enters a Coulomb-dominated regime ($\Ec \gg \J$); its eigenstates evolve toward charge eigenstates with energies approximately described by the charging energy term.
The triple-junction maintains its $\mathbb{Z}_3$ symmetry since the total number of Cooper pairs is conserved modulo 3. However, it cannot be described any longer as a clock degree of freedom.

\section{The Josephson junction chain}\label{sec:JJA}
To design a system with a global $\mathbb{Z}_3$ symmetry, we consider a chain of SC islands connected to the same background superconductor via the triple Josephson junctions described in the previous section (see Fig.~\ref{fig_device}), which provides a specific example of the models presented in Ref. \cite{Roy2023.2}. 
Each SC island is coupled with its neighbors through a standard sinusoidal Josephson junction of energy $E_J$ and the related Hamiltonian reads:
\begin{multline} \label{hamtot}
H= -\sum_{j=1}^L \J \cos 3 \varphi_j -\sum_{j=1}^{L-1} E_J \cos \left(\varphi_j - \varphi_{j+1} + A_j \right) \\ + \sum_{j=1}^L \Ec \left(N_j - n_g\right)^2\,,
\end{multline}
where $A_j$ labels the Peierls phase acquired by a Cooper pair when tunneling from the $j^{\rm th}$ to the $(j+1)^{\rm th}$ island. This system, depicted in Fig.~\ref{fig_device}, is indeed subject to a uniform out-of-plane magnetic field adjusted to reproduce a flux $\Phi_0/3$ in each of the SC loops within any of the triple junctions. The loop defined by neighboring SC islands may instead be characterized by a different area, thus resulting in a different Aharonov-Bohm phase when a Cooper pair encircles it. These Aharonov-Bohm phases, which we label as $2\pi\Phi_j$, depend on the phases $A_j$ in Eq. \eqref{hamtot} via the equivalence $2\pi\Phi_j = -4\pi/3 - A_j$.
If the system geometry is such that all the fluxes $\Phi_j=\Phi_{\rm l}$ are equal, also the phases $A_j$ become position independent and result:
\begin{equation}
A= 2\pi\Phi_{\rm l}+ \frac{4\pi}{3}\,.
\end{equation}
The introduction of Josephson junctions between neighboring islands also favors the onset of mutual electrostatic interactions between the islands of the form $4e^2( C^{-1})_{j,j'}\left(N_j-n_g\right)\left(N_{j'}-n_g\right)$, where $(C^{-1})_{j,j'}$ are the off-diagonal elements of the inverse capacitance matrix of the system \cite{Fazio2001rev}. In Eq. \eqref{hamtot} these interactions are neglected: we expect indeed that the self-capacitance between each island and the grounded SC background is considerably stronger than the mutual capacitance between the islands. In particular, Ref. \cite{Materise2023} investigated a setup with two transmons connected to the same SC background and linked by a tunable hybrid Josephson junction. In that analysis the maximal ratio between the diagonal terms $(C^{-1})_{j,j}$ and the off-diagonal nearest-neighbor term $(C^{-1})_{j,j+1}$ was about $1/9$ and it was strongly suppressed when increasing the filling the semiconducting region between the two neighboring superconducting islands as effect of the  screening provided by the two-dimensional electron gas. The role of inter-island interactions, however, does not qualitatively affect the main features of the system and we will discuss it in Sec. \ref{sec:pert_dmrg}.

The Hamiltonian \eqref{hamtot} is characterized by a global $\mathbb{Z}_3$ symmetry that corresponds to a simultaneous phase slip of all the SC islands: $\varphi_j \to \varphi_j + 2\pi/3$ for each $j$. This discrete symmetry characterizes the Josephson junction chain when all the triple Josephson junctions are tuned to display a potential of the form \eqref{pot}; the symmetry is independent on the values of $\J, E_J, \Ec, n_g$ and the phases $A_j$, and it is not affected by disorder in these parameters. 

The role of this discrete symmetry becomes more explicit when considering the low-energy dynamics of the system in the transmon-like regime with $\Ec \ll \J$ and $E_J \ll \sqrt{\J \Ec}$. In this high-$\J$ regime, each SC island can be described through the semiclassical approximation discussed in Sec. \ref{sec:TJJ} and the interaction $E_J$ between neighboring island is weaker than the gap separating the three low-energy states $\ket{s}$ of each building block from higher energy excitations. Therefore we can treat each island as a 3-level quantum clock, and their low-energy many-body dynamics is appropriately described by the following chiral quantum clock Hamiltonian \cite{fendley2012,Howes1983}:
\begin{multline} \label{Potts}
H = -\frac{E_J}{2}\sum_{j=1}^{L-1} \left[\ee^{i A_j} \sigma^\dag_{j+1} \sigma_j + \rm{H.c.}\right] \\
- h_\tau \sum_{j=1}^{L} \left(\ee^{-i n_g \frac{2\pi}{3}} \tau +  \ee^{i n_g \frac {2\pi}{3}} \tau^\dag\right)\,.
\end{multline}
Here the operators $\sigma_j$ and $\tau_j$ are clock operators referring to the SC island $j$; in particular, the operator $\sigma_j \sim \ee^{i\varphi_j}$ constitutes a discretized approximation of the annihilation operator for the Cooper pairs in the island $j$. The phases $A_j$ break time-reversal $(\sigma \to \sigma^\dag)$ and space-inversion $(j\to L+1-j)$ symmetries, therefore, in general, the Hamiltonian \eqref{Potts} represents a chiral clock model. In particular, the $E_J$ term represents a ferromagnetic interaction for $|A_j| < \pi/3$; the parameter $h_j$, instead, can be estimated via Eq. \eqref{slip} and it corresponds to a transverse magnetic field for the clock model. The induced charge $n_g$, instead, breaks the symmetry under particle-hole conjugation ($\sigma \to \sigma^\dag$ and $\tau \to \tau^\dag$) and plays the same role as the phases $A_j$ under Kramers-Wannier duality \cite{Whitsitt2018}.

The global $\mathbb{Z}_3$ symmetry corresponds to the operator $P=\prod_j \tau_j= \ee^{i\frac{2\pi}{3}N_{\rm tot}}$, with $N_{\rm tot} = \sum_j N_j$ being the operator that defines the total number of Cooper pairs, which is conserved modulo 3.

Finally, we observe that in uniform JJAs with position-independent phases $A$, both Hamiltonians \eqref{hamtot} and \eqref{Potts} fulfill the following properties: 
\begin{itemize}
\item the space inversion symmetry $j \to L+1-j$ maps $H(A) \to H(-A)$;
\item the gauge transformation $\varphi_j \to \varphi_{j} + 2\pi j/3$ maps $H(A) \to H(A-2\pi/3)$.
\end{itemize}
Therefore, in the following we can restrict our analysis to the range $0\le A \le 2\pi/3$ and the phase diagram of the system must be symmetric under $A \to 2\pi/3 -A$.

\section{Effective field theory description of the model}\label{sec:ft}

The Hamiltonian \eqref{hamtot} yields a rich phase diagram as a function of its parameters.
To obtain a low-energy description of this JJA, it is convenient to introduce an effective field theory that fulfills the same discrete symmetries and corresponds to the continuum limit of Eq.~\eqref{hamtot}. To this purpose we apply a bosonization approach \cite{Glazman1997,Giamarchi2003,maffi2023}: we promote the discrete operators $\varphi_j$ to a one-dimensional quantum field $\varphi$ and we introduce its dual field $\theta$, such that $\left[\theta(y),\varphi(x)\right]=-i\pi\Theta(y-x)$, where $\Theta$ is the Heaviside step function. Consequently, the charge operator $N_j$ corresponds to $-a\partial_x\theta/\pi$, with $a$ being the lattice spacing of the Josephson junction chain. In the harmonic approximation for the Josephson coupling along the chain, the dynamics of the model is approximated by:
\begin{multline} \label{LL}
H= \frac{v}{2\pi}\int dx\left[ K \left(\partial_x \varphi -A\right)^2 + \frac{1}{K}\left(\partial_x \theta - \frac{\pi n_g}{a}\right)^2 \right]\\
- \int dx\left[ \frac{\J}{a} \cos 3\varphi + M \cos 2\theta\right].
\end{multline}
This description corresponds to a dual sine-Gordon model with $\mathbb{Z}_3$ symmetry (see, for example, \cite{Lecheminant2002,Matsuo_2006}).
The first two terms respectively correspond to the Josephson coupling $E_J$ between neighboring islands and the charging energy of each island; they define a Luttinger liquid with Luttinger parameter and velocity approximately given by \cite{maffi2023,Glazman1997}:
\begin{equation}\label{eq:luttinger_p}
K = \pi \sqrt{\frac{E_J}{2\Ec}}\,,\qquad v=a\sqrt{2E_J\Ec}\,.
\end{equation}
In systems with sizeable mutual interactions between neighboring islands, these electrostatic repulsions further enhance the role of $E_c$, thus lowering $K$ and increasing $v$.

The phase $A$ and induced charge $n_g$ in Eq. \eqref{LL} constitute two incommensuration parameters: they respectively favor states with windings of the fields $\varphi$ and $\theta$ which, in general, are not commensurate with the lattice spacing. Their effect can yield the onset of Lifshitz transitions and floating gapless phases \cite{Ostlund1981,Howes1983,Milsted2014,Whitsitt2018} that compete with the gapped phases arising when either of the sine-Gordon terms dominates. 

Concerning the sine-Gordon interactions, the $\J$ term directly models the triple-junction Josephson potential and favors superconducting ground states in which the $\varphi$ phase is ordered and pinned to the minima of $V(\varphi)$. This sine-Gordon term is responsible for the breaking of the U(1) symmetry typical of Luttinger liquids into the $\mathbb{Z}_3$ symmetry corresponding to $\varphi\to \varphi+2\pi/3$. 

The $\cos(2\theta)$ term, instead, corresponds to backscattering processes of the Cooper pairs originating from the charging energies of the islands. This potential favors states in which the $\theta$ field is pinned, which correspond to the insulating phases typical of the Coulomb-dominated regimes, in which the field $\varphi$ becomes disordered. The parameter $M$ can be estimated in the regime $\Ec \gg \J,E_J$, where the $\cos(2\theta)$ term is the most relevant. This interaction opens a gap, which, on one side, corresponds to the energy $\Ec (1-2n_g)$ of a charge excitation in the Coulomb-dominated regime; on the other, it must match the mass of the related kink $M_{\rm kink}=4\sqrt{Mv/\pi K} -2\Ec n_g$ \cite{Nyhegn2021}. By equating them, we obtain:
\begin{equation} \label{Mparameter}
M= \frac{\pi^2}{32a}\Ec\,,
\end{equation}
see further details in Appendix \ref{sec:kink}.

Additional electrostatic interactions among neighboring islands would influence the parameters $K,v$, and $M$ without qualitatively affecting the field Hamiltonian \eqref{LL}.

This low-energy description of the system allows us to gain important insights into the order parameters and correlations that help identifying the onset of the zero-temperature phases of the Josephson junction chain. In particular, in the next sections we will investigate the interplay between charging energy, Josephson interactions, and the incommensurability parameters $A$ and $n_g$, which yields a rich phase diagram in the same universality class of the chiral clock model \cite{Howes1983}.

\section{Observables and phase diagram}\label{sec:results}

The phase diagram of the Josephson junction chain emerging from the Hamiltonian \eqref{hamtot} presents three main types of phases: (i) gapped ordered phases characterized by a spontaneous breaking of the $\mathbb{Z}_3$ symmetry; these correspond to superconducting phases dominated by the Josephson coupling $E_J$. (ii) gapped disordered phases, which correspond instead to insulating states dominated by the charging energy $\Ec$. (iii) a gapless Luttinger phase emerging due to the effect of the incommensuration parameters $A$ and $n_g$; this critical phase is characterized by a superconducting quasi-long-range order.

These phases have precise counterparts in the field theory description of Eq. \eqref{LL}. The ordered phases correspond to regimes in which the $\cos 3 \varphi$ is the most relevant and the superconducting phase acquires long-range order. The disordered phases correspond to Mott-insulating regimes in which, instead, the $\cos 2 \theta$ term becomes the most relevant and the charge fluctuations are suppressed. Finally, gapless phases appear when both these interactions are irrelevant, and all the two-point correlation functions display a power-law decay.

To explore these phases and study their correspondence with the $\mathbb{Z}_3$ quantum clock model in Eq. \eqref{ham2}, we perform an extensive DMRG analysis aiming at characterizing the ground state properties in the different regimes and estimating the related correlations and order parameters. To this purpose, we use the charge basis and set a charge cutoff $|N|\le N_{\rm max}=7$ for each island, which consists of a local Hilbert space dimension 15. 
In particular, in our simulations, we consider charging energies $\Ec$ sufficiently large to avoid strong cutoff effects. To this purpose, we check that the population of the charge states at $N=\pm N_{\rm max}$ is always below $10^{-3}$.

In this truncated space, the creation and annihilation operators of the Cooper pairs, $\ee^{-i\varphi_j}$ and $\ee^{i\varphi_j}$, are respectively approximated by off-diagonal $(2N_{\rm max}+1) \times (2N_{\rm max}+1)$ matrices $\Sigma_j^+$ and $\Sigma_j^-$, which raise and lower the particle number of the island $j$ (see, for instance, Ref. \cite{maffi2023}). We also observe that, in the large $\J$ limit, where the mapping to the clock model \eqref{ham2} is most accurate, the matrices $\Sigma_j^+$ and $\Sigma_j^-$ respectively converge to the clock operators $\sigma^\dag_j$ and $\sigma_j$. Therefore the correspondence between phase operators, truncated matrices for their description, and clock operators is given by:
\begin{align}
&\ee^{i\varphi_j} \approx \Sigma^-_j \sim \sigma_j\,,\\
&\ee^{-i\varphi_j} \approx \Sigma^+_j \sim \sigma^\dag_j\,,\\
&\ee^{i\frac{2\pi}{3}N_j} \sim \tau_j\,.
\end{align}

Each SC island is represented as a site of a matrix product state (MPS) with open boundaries and maximum bond dimension $\chi=800$. 
During the DMRG optimization, we monitor the convergence of the variational energy and stop the iterations once a precision of $10^{-6}$ has been reached. 
The accuracy of the MPS representation is set to $\epsilon = 10^{-10}$, which determines the scale of the smallest singular value we retain in the SVD truncation. 
Our simulations are performed with the ITensor julia library~\cite{Fishmann2022,itensor-r0.3}.

\subsection{The Potts model regime: $n_g=0$, $A=0$} \label{sec:potts}

We begin our investigation by targeting the physics of the achiral clock model in a scenario that preserves the inversion and time-reversal symmetries. This case is realized by setting $n_g=0$ and $A_j=0$ in Eq.~\eqref{Potts}, such that this effective description of the system in the high-$\J$ limit becomes equivalent to the 3-state Potts model. 

We set, in particular, $E_J=0.2\J$ and vary $\Ec/E_J$ to study the expected Potts phase transition. For low $\Ec$ the system lies in its symmetry-broken and ordered phase. This corresponds to the ferromagnetic phase of the clock model which is characterized by the following two-point correlation function converging exponentially towards a constant at large distances $j$:
\begin{equation} \label{corr}
\left\langle \ee^{i\varphi_{j_0}} \ee^{-i\varphi_{j_0+j}} \right\rangle\approx m^2 + \delta m^2 \,\ee^{-j/\xi}, \; \text{for} \; j\gg 1\,.  
\end{equation}
Here we introduced the effective order parameter $m$, which corresponds to the superconducting order parameter for the Josephson junction chain and plays the role of the magnetization of the clock model. The parameter $\xi$ constitutes the related correlation length.

\begin{figure}
    \centering
    \includegraphics[width=\columnwidth]{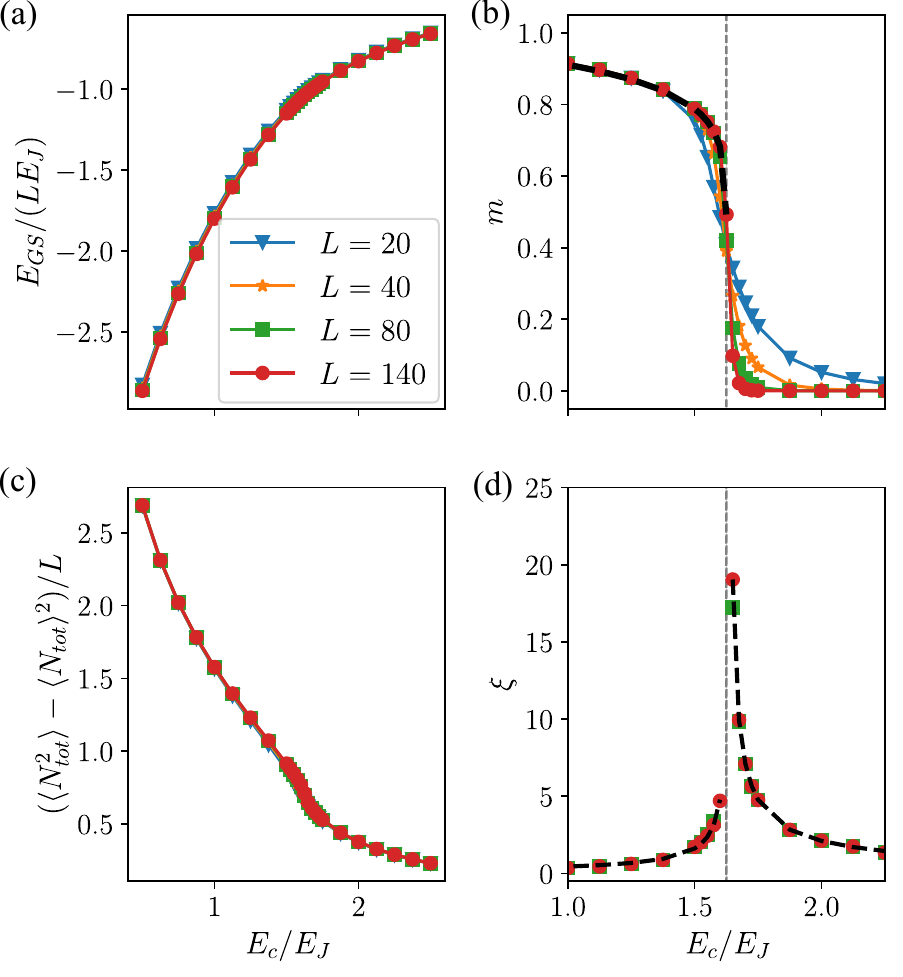}
    \caption{Summary of results obtained with DMRG: ground state energy, long-range order parameter, total charge fluctuations, correlation length. The data display the comparison between four system sizes $L$ with a charge cutoff of $N_{\rm max}=7$, corresponding to 15 different charge states for each SC island.
    The other physical parameters are $E_J=0.2\J$, $n_g=0$, and $A =0$. 
    The vertical dashed lines in panels (b) and (d) mark the phase transition occurring at $\Ec=\Ec^*\simeq 1.62 E_J$. The dashed black lines are power-law fits returning critical exponents $\beta=0.089(2)$, $\nu^+=0.88(5)$ for $\Ec>\Ec^*$ and $\nu^-=0.87(6)$ for $\Ec<\Ec^*$.}
    \label{fig:results_phi0}
\end{figure}

Our DMRG findings are summarized in Fig.~\ref{fig:results_phi0}, where we show the ground state energy density of the system (a), the order parameter $m$ (b), the charge fluctuations (c), and the correlation length $\xi$ (d) for different lengths $L$ of the chain. 

Both the ground state energy and the charge fluctuation per lattice site, panels (a) and (c), display very little dependence in the length $L$, ensuring that the simulations always converge to a physical state where there are no sizable finite-size effects.
As expected, the charge fluctuations increase for decreasing charging energy. This corresponds to a transition between a Mott insulating state at large $\Ec$ to the superconducting phase for low $\Ec$. In particular, in the $\Ec\to 0$ limit the JJA becomes grounded. We observe, however, that the lowest value of $\Ec$ that we considered for the data in Fig. \ref{fig:results_phi0} is $\Ec =E_J/2=0.1\J$ such that the corresponding charge fluctuation per island is still considerably smaller than the cutoff, making us confident that the representation we use is accurate.

\begin{figure}
    \centering
    \includegraphics[width=\columnwidth]{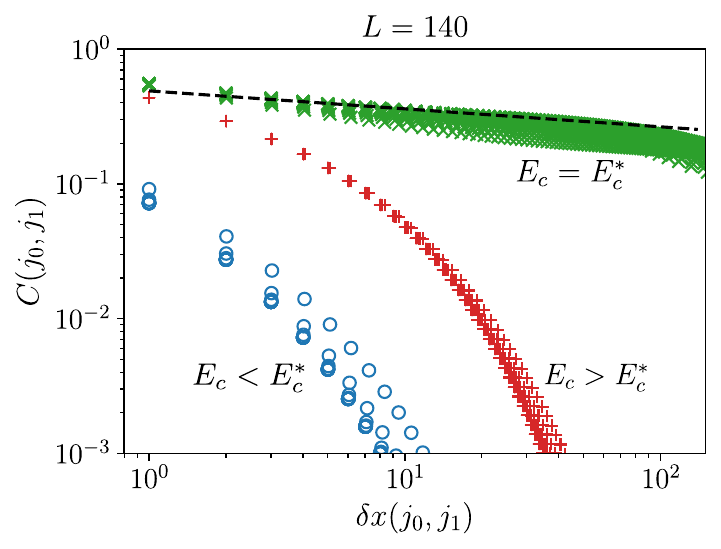}
    \caption{Connected correlation function $C(j_0,j)$ (Eq. \eqref{corr}) vs the chord distance $d(j_0,j)$ (Eq. \eqref{eq:chord_mod}) for different values of $j_0\in[10,L/2]$ and $j_1 \in[j_0, L-10]$ at $\Ec = \Ec^*$ (green crosses). We excluded the first and last ten sites to avoid boundary effects. The value of $\Ec^*$ is derived from the fits in Fig. \ref{fig:results_phi0}. The black dashed line highlights the decay $C(L/2,x)\propto \delta x^{-D_\sigma}$, with $D_\sigma =2/15$ marking the Potts critical behavior.
    The red crosses and blue dots display the typical exponential decay in the disordered and ordered phases respectively. 
    }
    \label{fig:corr}
\end{figure}

The correlation function \eqref{corr} is calculated via the approximation:
\begin{equation}
\tilde{C}(j_0,j)= \langle\Sigma^+_{j_0}\Sigma^-_{j_0+j}\rangle\,.
\end{equation}
Correspondingly, we estimate the order parameter as  $m=\sqrt{\tilde{C}(j_0,j_0+L/2)}$ averaged over $L/4$ different reference positions $j_0<L/2$.
The onset of a second-order phase transition is evident from its behavior in Fig. \ref{fig:results_phi0} (b):
as the system size increases, $m$ drops to zero more and more sharply for large $\Ec$, whereas it converges to finite values for low $\Ec$. Curves related to different system sizes $L$ all cross at the same point, allowing us to estimate the critical charging energy $E^*_c=1.62E_J$.
A power-law fit $m\propto (\Ec^*-\Ec)^\beta$ of the data with $L=140$ for $\Ec<\Ec^*$ (dashed black line) provides the critical exponent $\beta=0.089(2)$ which is in good agreement with the expected universal critical exponent for the Potts phase transition $\beta = 1/9$ \cite{mussardobook,difrancesco}.
This agreement is further confirmed by the correlation length $\xi$, shown in panel (d), which we extract through an exponential fit of the connected correlation function:
\begin{equation}
C(j_0,j)=\langle\Sigma^+_{j_0}\Sigma^-_{j_0+j}\rangle-\langle\Sigma^+_{j_0}\rangle \langle\Sigma^-_{j_0+j}\rangle\,.
\end{equation}
$\xi$ has a power-law divergence at the quantum critical point $\Ec^*$ with exponents $\nu^+=0.88(5)$ ($\Ec > \Ec^*$) and $\nu^-=0.87(6)$ ($\Ec<\Ec^*$).
The Potts phase transition predicts a symmetric critical exponent $\nu=5/6$ which is compatible with our numerical results.

urther confirmations of the correspondence between the critical behavior of the JJA and the $\mathbb{Z}_3$ clock model emerge from the analysis of the power-law decay of $C(L/2,j)$ at the critical point.
Figure \ref{fig:corr}(a) illustrates the algebraic decay of the connected correlation function at the critical point (green crosses), as a function of the renormalized chord distance \cite{Cazalilla2004}
\begin{equation}\label{eq:chord_mod}
    d(j_0,j) = \frac{c(j_0-j,2L)c(j_0+j,2L)}{\sqrt{c(2j_0,2L)c(2j,2L)}} \ ,
\end{equation}
where $c(j,L)=\frac{L}{\pi}\left|\sin \left( \frac{\pi j}{L} \right)\right|$ labels the standard chord distance. This choice of the distance $d(j_0,j)$ reflects open boundaries in which we assume Neumann boundary conditions for the field $\varphi$ \cite{Cazalilla2004,Haller2020}. This condition is not exactly fulfilled by the JJA, but it provides a rough approximation that allows us to limit the boundary effects.

The correlation function $C(j_0,j)$ at $\Ec = \Ec^*$ displays a power-law decay with a critical exponent compatible with the Potts universality class $D_\sigma = 2/15$ (dashed black line). For values of $\Ec$ lying in the two gapped phases, instead, the connected correlation function $C$ decays exponentially with the distance as expected (blue circles and red crosses).

\subsection{The chiral classical limit: $\Ec \to 0$, $A \neq 0$} \label{sec:class}

The vector potential $A$ in the Hamiltonian \eqref{hamtot} breaks the time-reversal and space-inversion symmetries and it causes a competition between the onsite potential \eqref{pot} and the nearest-neighbor Josephson interaction $E_J$. These two terms cannot indeed be simultaneously minimized if $A$ is not a multiple of $2\pi/3$.

To understand their competition, we focus first on the classical limit of the Hamiltonian \eqref{hamtot}, $\Ec=0$, which corresponds to a one-dimensional rotor model and displays three kinds of classical ground states, depending on $A$ and the ratio $\J/E_J$. 

For $\J/E_J>1/2$, the classical limit of the Josephson Hamiltonian \eqref{hamtot} converges to the classical limit of the chiral clock model \eqref{Potts} ($h_\tau=0$). This regime is characterized exclusively by ordered phases in which the clock degrees of freedom $\sigma_j$ are either aligned in a ferromagnetic state (if $|A| < \pi/3$) or wind as $\sigma_j = \ee^{i2\pi w j/3}$ with winding $w=1$ for $\pi/3< A <\pi$ or $w=-1$ for $\pi< A <5\pi/3$. These two states with $w\neq 0$ correspond to helical phases in which the system displays long-range order but acquires a chiral behavior.
If $\J/E_J<1/2$, instead, a third kind of ground state emerges in which the winding of the rotor is not commensurate with the lattice spacing, and the model cannot be simply described in terms of the clock limit. In particular, ground states with a phase winding given by $\varphi_j = Aj$ (up to an overall shift) compete with the previous commensurate configurations.

\begin{figure}
    \centering
    \includegraphics[width=\linewidth]{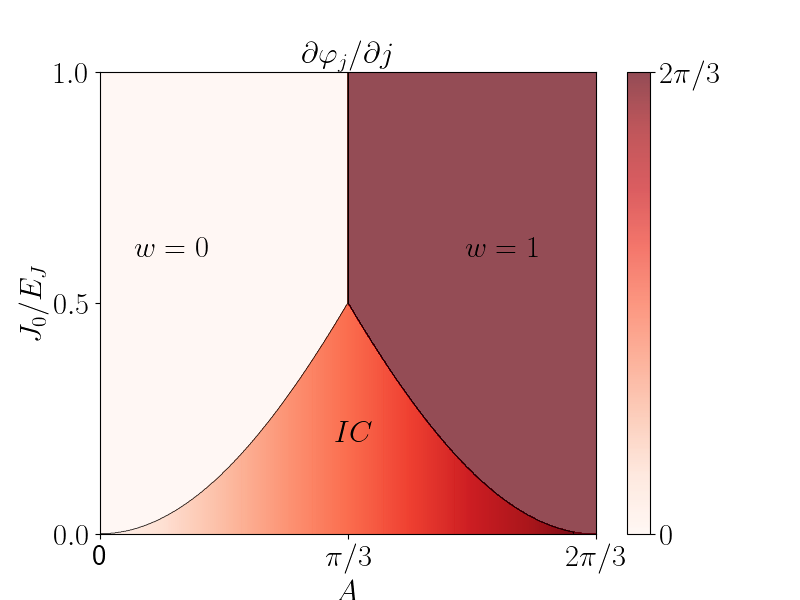}
    \caption{Phase diagram of the classical limit ($\Ec=0$) of the Josephson junction chain. Depending on $A$ and the ratio $J_0/E_J$ the classical ground state corresponds to the ferromagnetic phase ($w=0$), the helical phase ($w=1$), or the incommensurate (IC) phase ($\varphi_j=Aj$).}
    \label{fig:class} 
\end{figure}

Given the symmetries of the system, we restrict our analysis to $0\le A \le 2\pi/3$. The classical configurations with minimal energy are depicted in Fig.~\ref{fig:class} and the boundaries between the commensurate and the incommensurate classical ground states are easily found by comparing their average energy densities: the energy density of the ferromagnetic configuration is given by $\varepsilon= -\J -E_J\cos A$; the one for the incommensurate phase with winding $A$ is instead $-E_J$, since the onsite potential averages to zero. Their boundary is thus given by:
\begin{equation} \label{arccos}
A=\arccos \frac{E_J-\J}{E_J}\,.
\end{equation}
The boundary between the helical phase with $w=1$ and the incommensurate phase is obtained in an analogous way (Fig. \ref{fig:class}). 

For $\J/E_J>1/2$, there is a direct transition between the ferromagnetic and the chiral ordered phases at $A=\pi/3$. The transition between them, however, is not trivial and can be understood in terms of the clock model. For $A=\pi/3$, the classical system displays two degenerate configurations with the lowest energy for each link between neighboring clock degrees of freedom: either $\sigma_{j+1}=\sigma_j$ or $ \sigma_{j+1}= \ee^{i2\pi/3} \sigma_j$. We can thus introduce the notation $\sigma_{j+1}=\ee^{i2\pi w_j/3} \sigma_j$ with the link variable $w_j=0,1$. Given the energy degeneracy of links with $w_j=0,1$, the classical system acquires an extensive entropy \cite{Berker1980,Howes1983}, with a number of classical ground states scaling exponentially as $2^{L-1}$ with the system size.

This exponential degeneracy is split when leaving the classical limit by introducing a weak charging energy $\Ec>0$, thus $h_\tau>0$. The behavior of the JJA for $A\sim \pi/3$ and $\J>E_J/2$ can be described based on a mapping of the chiral clock into a system of free fermions presented by Ostlund \cite{Ostlund1981}. The link variable $w_j=0,1$ can indeed be considered as the occupation number of a hardcore boson since link configurations of the clock model with $w_j=2$ are suppressed by an energy gap $3E_J/2$, playing the role of an onsite interaction term of the bosonic degrees of freedom. These hardcore bosons can be mapped into fermionic degrees of freedom by a Jordan-Wigner transformation. Additionally, the transverse field term $h_\tau$ in Eq. \eqref{Potts} becomes a kinetic energy term for these fermions. The model for $A=\pi/3$, $\J>E_J/2$ and $\Ec \ll E_J$ can thus be mapped into a system of fermions at half filling with bandwidth $2h_\tau$.

If the phase $A$ is slightly displaced from $\pi/3$, the links with $w_j=0,1$ acquire a small energy shift:
\begin{equation} \label{muost}
\mu= -\sqrt{3}E_J\sin\left( A-\frac{\pi}{3}\right).
\end{equation}
This shift plays the role of a chemical potential for the hardcore boson description.
Finally, for weak $h_\tau \ll E_J$, at second order in perturbation theory, a further term emerges that can annihilate three neighboring bosons $w_j=1$ by rotating two rotors. Overall, the model at low $\Ec$ in proximity of $A=\pi/3$ can thus be described by:
\begin{multline} \label{hambos}
H_{b} = \mu \sum_{i} w_i -h_\tau\sum_i\left( b^\dag_{i+1}b_i + {\rm H.c.}\right) \\
+\frac{3E_J}{4}\sum_i w_i \left(w_i-1\right) -\frac{4h_\tau^2}{3E_J} \sum_i \left( b_{i+1}b_ib_{i-1} + {\rm H.c.}\right)\,,
\end{multline}
where $w_i=b^\dag_i b_i=0,1,2$ labels the twist of the phase $\varphi$ in units of $2\pi/3$ across the $i^{\rm th}$ link of the clock model, and the operators $b_i,b^\dag_i$ represent bosonic annihilation and creation operators in the corresponding bosonic description. 
The Hamiltonian \eqref{hambos} can also be interpreted under the point of view of the field theory description \eqref{LL}. 
In particular, the last term of Eq. \eqref{hambos} in the continuum limit corresponds to the $\cos 2 \theta$ sine-Gordon interaction in Eq. \eqref{LL}. 
This interaction is highly irrelevant in the limit of small $\Ec$, therefore it does not open a gap close to the classical regime. For this reason, Eq. \eqref{hambos} describes a gapless phase for low $h_\tau$ which opens from the degenerate classical states at $A=\pi/3$ when including a weak $\Ec$. 
In Fig. \ref{fig:phase_diag} we show the prediction obtained by setting $|\mu|=h_\tau$ about the opening of this gapless phase for $\Ec>0$ (black lines). 
This prediction, however, is accurate only in the limit $\Ec \ll E_J$ and considerable deviations are expected for increasing $\Ec$ due to corrections both in the perturbative description \eqref{hambos} and the validity of the semiclassical approximation \eqref{slip}.

\subsection{Chiral model at $n_g=0$ and the floating phase}\label{sec:chiral}

A more extensive analysis of the model for generic $\Ec$ and $A$ can be performed based on the field theory description \eqref{LL} and DMRG simulations.
For the sake of simplicity, we restrict here to a single incommensuration parameter by setting $n_g=0$.

\begin{figure}
    \centering
    \includegraphics[width=\columnwidth]{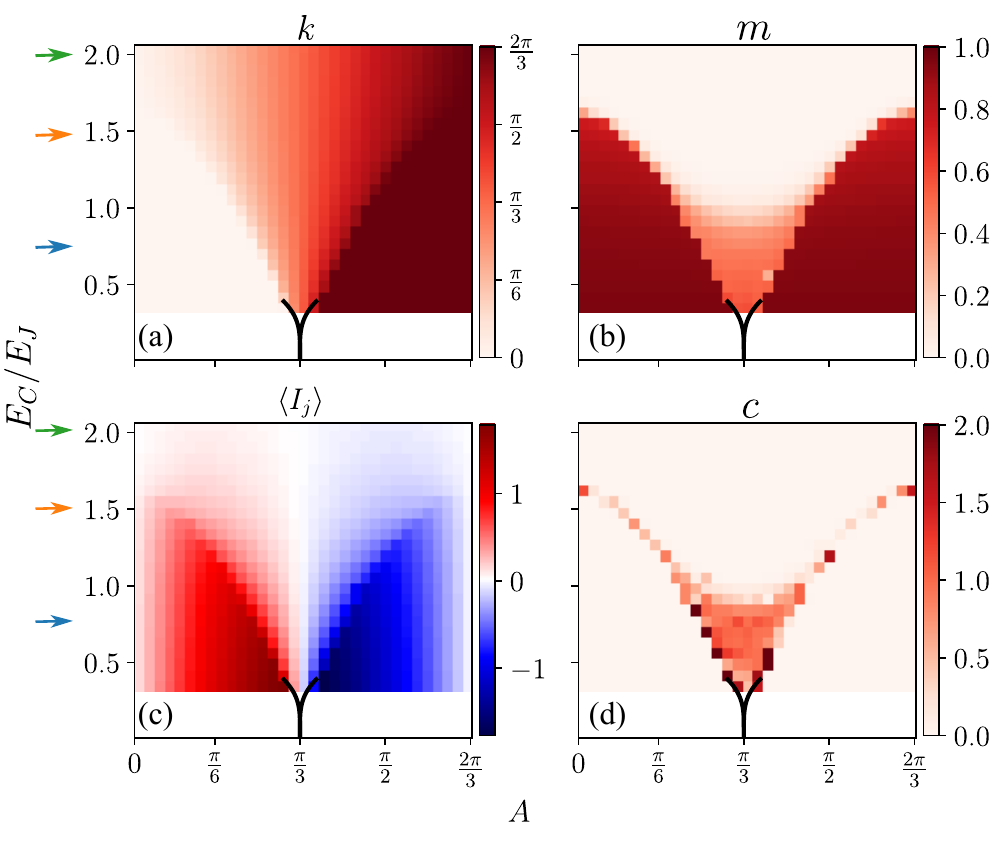}
    \caption{Summary of the phase diagram in the $\Ec-A$ plane for $\J = 5E_J$:
    (a) phase winding $k$, (b) order parameter, (c) current density, (d) central charge.
    The continuous black lines mark the analytical estimate of the phase transitions at small charging energies, where the charge truncation would induce nonphysical effects in our MPS simulations.
    The colored arrows indicate the horizontal cuts of the phase diagram used for Figs.~\ref{fig:flux} and \ref{fig:current}.
    The system size is $L=100$.}
    \label{fig:phase_diag}
\end{figure}

The classical configurations depicted in Fig. \ref{fig:class} clearly show that we must distinguish two regimes for $J_0 \gtrless E_J/2$. The Josephson junction system for $J_0>E_J/2$ converges to the chiral clock model \eqref{Potts} for $J_0 \to \infty$. In this regime, the phase diagram of the Josephson chain is exemplified by Fig. \ref{fig:phase_diag} and it qualitatively matches the corresponding phase diagram for a 2D classical clock model at finite temperature analyzed by Ostlund in Ref.~\cite{Ostlund1981} and its quantum counterpart in Refs.~\cite{Howes1983,Zhuang2015,Samajdar2018}. In particular, as seen in Fig.~\ref{fig:phase_diag}, the classical states with extensive entropy at $\Ec=0$ and $A=\pi/3$ become a gapless phase at finite $\Ec$ that can be understood both in terms of the gapless phase of the model \eqref{hambos} and in terms of the Luttinger liquid description \eqref{LL} when $A$ is strong enough for the potential $\cos3\varphi$ to be irrelevant.

For $\J<E_J/2$ (see Appendix \ref{app:extradata_phasediagr} and Fig. \ref{fig:phase_diag2}), instead, the gapless Luttinger phase extends for a finite interval of the parameter $A$ also in the classical limit $\Ec\to 0$ and it corresponds to the extension of the incommensurate classical states in Fig. \ref{fig:class}.  

This Luttinger phase is conveniently characterized by the correlation function \eqref{corr} which, on one hand, shows a power-law decay and, on the other, displays an incommensurate winding. By considering the continuum and thermodynamic limit, the field theory \eqref{LL} predicts behavior of the kind
\begin{equation}
C(x,y) \propto \frac{\ee^{-i k (x-y)}}{|x-y|^{\frac{1}{2K}}}\,.
\end{equation}
Given its non-commensurate wave parameter $k$ and the gapless nature of this phase, we refer to this Luttinger phase as \textit{floating}.
In particular, the wave parameter $k$ becomes $\pi/3$ for $A=\pi/3$ and, for the case $\J<E_J/2$ it converges to $A$ in proximity of the incommensurate classical phase.

The incommensurate winding of the correlation function $C$ is not limited to the gapless phase, but it also affects the paramagnetic disordered phase that characterizes the phase diagram (Figures \ref{fig:phase_diag} and \ref{fig:phase_diag2}) at high values of $\Ec$, where $C$ decays exponentially with the distance:
\begin{equation}
C(x,y) \propto \ee^{-i k (x-y) - \frac{|x-y|}{\xi}}\,.
\end{equation}

\begin{figure}
    \centering
    \includegraphics[width=\columnwidth]{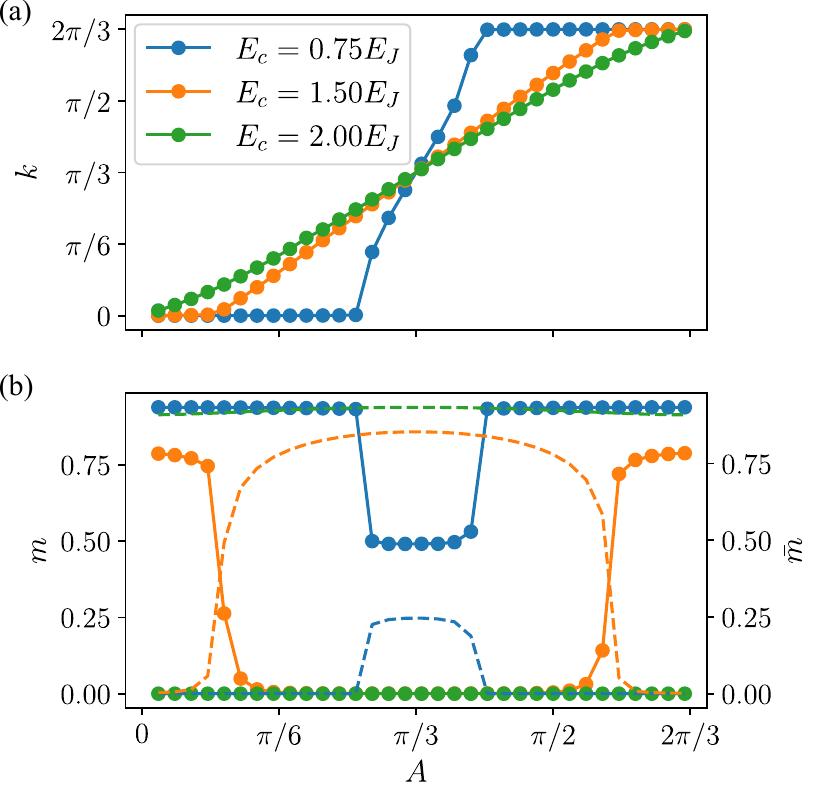}
    \caption{(a) Phase winding and (b) order parameter vs the Aharonov-Bohm phase $A$ for three different charging energies $\Ec$, $\J = 5E_J$, and $L=100$. In panel (b), the full circles represent the order parameter extracted from the correlation function $m=\sqrt{|\tilde{C}(j_0,j_0+L/2)|} $, while the dashed lines depict the string order parameter $\bar{m}=\sqrt{|\tilde{S}(j_0,j_0+L/2)|}$.
    Notice the complementary behaviour of $m$ and $\bar{m}.$}
    \label{fig:flux}
\end{figure}

The phase diagram \ref{fig:phase_diag} is conveniently analyzed by comparing three qualitatively different horizontal cuts in the plane determined by $A$ and $\Ec$ at fixed $\J$ and $E_J$ (see the arrows in Fig. \ref{fig:phase_diag}). In particular, the phases of the systems can be understood by considering the winding of the connected correlations $C$ in Fig.~\ref{fig:flux}(a) and the order parameter $m$ in Fig.~\ref{fig:flux}(b) (full circles). 

The paramagnetic and ferromagnetic phases of the Potts limit are stable under the introduction of a small flux parameter $A$. The introduction of this incommensurate parameter, though, restricts the ordered phase as in the case of the chiral clock model \cite{Howes1983,Zhuang2015,Samajdar2018}. 

The green curve in Fig. \ref{fig:flux}(a) shows that the paramagnetic phase immediately acquires a non-commensurate winding $k$ for $A\neq 0$. This winding converges to $A$ sufficiently deep in the gapped disordered phase.
When considering the evolution of the ferromagnetic phase as a function of $A$, there are two distinct cases that are separated by a limiting value of the charging energy $\tilde{E}_c<\Ec^*$.

For $\tilde{E}_c < \Ec <\Ec^*$ (orange curves in Fig. \ref{fig:flux}), the system undergoes a direct second-order phase transition from the ordered to the disordered phase. The results in Refs. \cite{Samajdar2018,Whitsitt2018} based on the clock model suggest that this phase transition is not conformal, as it is characterized by a dynamical critical exponent $z>1$. This is consistent with the onset of incommensurate values of the winding $k$ which breaks Lorentz invariance. By increasing further the flux parameter, for $\pi/3 < A < 2\pi/3$ the system undergoes a symmetric transition from the paramagnetic to the ordered helical phase with $k=2\pi/3$.

\begin{figure}
    \centering
    \includegraphics[width=\columnwidth]{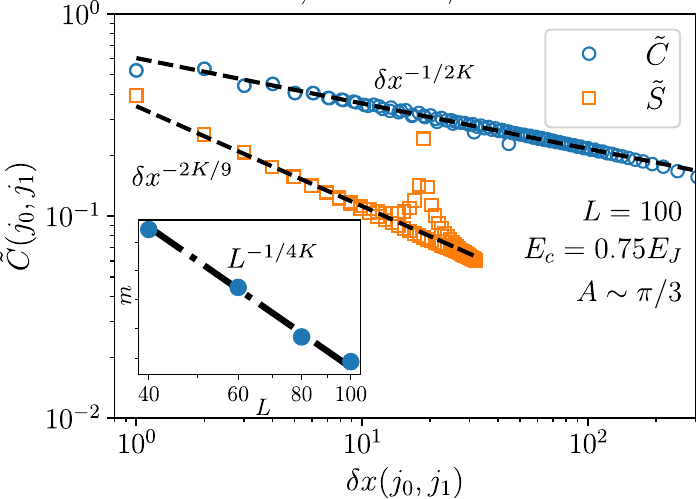}
    \caption{Decay of two-point correlation and string correlation as a function of the appropriate chord distance. For $\tilde{C}$, $\delta x(j_0,j_1)$ is given by $d(j_0,j_1)$ in Eq.~\eqref{eq:chord_mod}, whereas for $\tilde{S}$ it is the standard chord distance $c(j_1-j_0,L)$. The Luttinger parameter $K\sim 9/4$ is derived by a power-law fit of $\tilde{C}(j_0,j_1)$; the critical exponents of $\tilde{C}$ and $\tilde{S}$ satisfy the expected relations.
    For $\tilde{C}$ we show data corresponding to $j_0=10,20,\dots,50$, while for $\tilde{S}$ we depict only $j_0=20$ for clarity. For both $j_1 \in [1,L]$.
    The peak in $\tilde{S}$ is a boundary effect appearing when $j_1$ approaches either edge of the chain.
    Inset: $L^{-1 / 4K}$ decay of the order parameter with the system size in the floating phase.}
    \label{fig:correlation_K}
\end{figure}

For $\Ec < \tilde{\Ec}$, instead, the gapped ordered and disordered phases are separated by the gapless floating phase. In this case, the transition from the ordered to the floating phase is a commensurate-incommensurate Pokrosky - Talapov phase transition \cite{Pokrovsky1979,Nersesyan1979}. Besides the continuous change in the winding $k$ [Figures \ref{fig:phase_diag}(a) and \ref{fig:flux}(a)], this phase transition is also reflected in the different behavior of the order parameter: $m$ is different from zero and independent on the system size in the ordered phases, whereas it decays as a power law $L^{-\frac{1}{4K}}$ with the system size in the floating phase (see the inset of Fig. \ref{fig:correlation_K}). This decay is reflected in the considerably smaller value acquired by $m$ for finite-size systems as illustrated by Figures \ref{fig:phase_diag}(b) and \ref{fig:flux}(b).

To further confirm the nature of the phases displayed in Figs. \ref{fig:phase_diag} and \ref{fig:phase_diag2}, we additionally consider the disorder parameter $\bar{m}$ obtained by the asymptotic large-distance behavior of the following string operators:
\begin{equation} \label{string}
\tilde{S}(j_0,j) \equiv \left\langle \ee^{i\frac{2\pi}{3}\sum_{r=j_0}^{j} N_r} \right\rangle \sim \left\langle \ee^{-i\frac{2}{3}(\theta(j)-\theta(j_0))}\right\rangle\,.
\end{equation}
In the ordered phases, $\tilde{S}$ vanishes exponentially as a function of the distance $|j-j_0|$; in the floating phase it decays as a power law with exponent $-2K/9$; in the paramagnetic phase, it converges exponentially to the constant $\bar{m}^2$. The behavior of $\tilde{S}$ is thus dual with respect to $\tilde{C}$. Figure \ref{fig:flux} (b) displays the value of $\bar{m}$ for the three different cuts (dashed lines): $\bar{m}$ is different from zero in the paramagnetic phase, it acquires a smaller value (which decreases with the system size) in the floating phase, and it vanishes in the symmetry-broken phases.

Additional indications on the nature of these phases are provided by the study of the central charge derived by the entanglement entropy. By considering a partition of the system at position $x$, the central charge can be estimated by fitting the entanglement entropy through the Calabrese and Cardy formula:
\begin{equation} \label{CalabreseCardy}
    S_{\rm ent}(x) = \frac{c}{6} \log \left[ \frac{L}{\pi} \sin \left( \frac{\pi x}{L}\right)\right] .
\end{equation}
Eq. \eqref{CalabreseCardy} correctly captures the behavior of the entanglement entropy within the floating phase, whereas the gapped phases display a typical area law of $S_{\rm ent}(x)$. Fig.~\ref{fig:phase_diag} shows indeed that the floating phase displays a sizeable central charge with values spread around the value $c=1$ predicted from both the Luttinger liquid description \eqref{LL} and the Ostlund limit \eqref{hambos}.

A precise estimate of the values of $c$ is beyond our numerical precision, due to the large Hilbert space necessary for the description of the JJA and the dependence of this parameter on all the eigenvalues of density matrix. Our results, however, clearly confirm the onset of an extended gapless phase compatible with the Luttinger liquid description \eqref{LL}.

Additionally, the peaks of the entanglement entropy $S_{\rm ent}(L/2)$ allow us to confirm the critical values of the charging energy separating ordered and disordered phases. We observe, however, that even in the Potts limit $A\to 0$, a precise determination of the central charge $c$ is extremely challenging for our numerics, given the remarkably strong dependence of this parameter from tiny variations in $\Ec$. It is indeed known also for the study of the chiral clock model that small deviations of the parameters away from the critical point have a major impact on the fitted $c$ parameter \cite{Zhuang2015}.

Concerning the transition from the floating phase to the paramagnetic phase, we observe that the winding $k$ does not display any discontinuity. Based on the field theory description \eqref{LL} this transition happens when the sine-Gordon potential $\cos(2\theta)$ becomes relevant ($K=2$). Since this potential is not affected by the incommensurability parameter $A$, we conclude that this transition is of the Berezinskii-Kosterlitz-Thouless (BKT) kind. Therefore, when considering a vertical cut of the phase diagrams~\ref{fig:phase_diag} and \ref{fig:phase_diag2} such that all three phases are present, the two gapped/gapless phase transitions are of different nature: the first, from an ordered to the floating phase, is a non-conformal commensurate-incommensurate transition; the second, from the floating to the paramagnetic phase, is instead of the BKT class. This situation is analogous to the study of Rydberg atom systems with $\mathbb{Z}_3$ symmetry \cite{Giudici2019}. The BKT nature of the second transition, however, is not stable under the introduction of the other incommensurability parameter $n_g$, which turns also this phase transition into the commensurate-incommensurate kind.

To get further insight into the floating phase, we analyze the spatial dependence of two-point ($\tilde{C}(j_0,j_1)$) and string ($\tilde{S}(j_0,j_1)$) correlation functions. 
In the gapless phase, they both decay as a power-law of an appropriately chosen chord distance $\delta x(j_1,j_2)$ with the exponents set by the Luttinger parameter K.
Specifically, $\tilde{C}(j_0,j_1)\propto \delta x(j_0,j_1)^{\frac{1}{2K}}$, with $\delta x(j_0,j_1)$ appropriately described by $d(j_0,j_1)$ in Eq.~\eqref{eq:chord_mod}; the decay of the string operator is instead $\tilde{S}(j_0,j_1)\propto \delta x(j_0,j_1)^{-\frac{2K}{9}}$ for a suitable $\delta x(j_0,j_1)$ that, in principle, depends on the ratio of $E_J$ and $J_0$. In Fig.~\ref{fig:correlation_K} we adopted the rough approximation given by the standard chord distance $\delta x(j_0,j_1)= c(j_0-j_1,L)$.

These algebraic decays are perfectly captured by the data shown in Fig.~\ref{fig:correlation_K}; indeed, the Luttinger parameter $K$ extracted from $\tilde{C}(j_0,j_1)$ through a power-law fit matches very well both the decay of $\tilde{S}$ and the dependence of the order parameter $m$ on the system size (inset).
The only discrepancy is the peak displayed by $\tilde{S}$ when either $j_0$ or $j_1$ approaches the boundary. This is due to our suboptimal choice of the chord distance $c(j_0-j_1,L)$ that does not fully capture the boundary effects and prevents the data from folding back on the $\delta x^{-2K/9}$ curve. $\tilde{S}$ appears indeed as a multivalued function and only one of its branches rigorously follows the predicted algebraic decay. The determination of the most suitable conformal transformation to obtain a generalized chord distance \cite{Cazalilla2004} to fit the data related to $\tilde{S}$ would require a more detailed study of boundary conditions of the field theory in Eq. \eqref{LL} (see, for instance, the scattering formalism in Ref. \cite{Mintchev2006}). The power law exponent associated with $\tilde{S}$, however, is rigorously determined from the behavior at intermediate distances.

The value of $K$ we extract from the data is $K=2.22 \pm 0.27$, consistent with the fact that the considered value of $\Ec$ is close to the BKT phase transition into the disordered phase, which coincides with $K=2$.
The related estimate from the approximated expression of the Luttinger parameter, Eq.~\eqref{eq:luttinger_p}, returns $K=2.57$.

\begin{figure}
    \centering
    \includegraphics[width=\columnwidth]{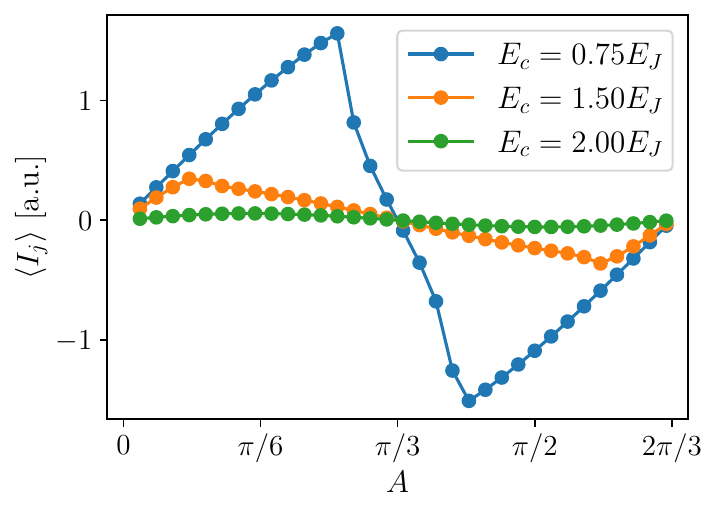}
    \caption{Current density as a function of the phase $A$ for three different charging energies. Here, $\J=5E_J$ and $L=100$, although the current per site displays very little or no size dependence.}
    \label{fig:current}
\end{figure}

After investigating the order and correlation properties of these phases, we consider their transport properties. The ordered phases are characterized by a superconducting long-range order. This is reflected into the distribution of the electrical current flowing across the JJA due to the introduction of the vector potential $A$:
\begin{equation}
I_j = E_J \sin \left(\varphi_j - \varphi_{j+1} + A \right).
\end{equation}
We show the expectation value of $I_j$ averaged over the position in Fig.~\ref{fig:current} which, as expected, is periodic in $A$ with a period $2\pi/3$.
The ordered phases are characterized by a Meissner current linear in $A$, whose appearance is made possible by the triple Josephson junctions connected with a superconducting background.
The commensurate-incommensurate phase transitions into the gapless phase are signaled by a sharp cusp in the current dependence on $A$, analogous to bosonic ladder models \cite{Orignac2001,Greschner2016}. The floating phase corresponds indeed to an incommensurate vortex phase and its average current per link decreases to zero for $A$ approaching $\pi/3$. Finally, the paramagnetic phase corresponds to a Mott insulator for the Cooper pairs. In this phase, the current is exponentially suppressed with the system size and shows little or no dependence on $A$.

\section{The effect of the induced charge}\label{sec:ng}

The induced charge $n_g$ must be considered an incommensurability parameter for the charge distribution $N$. In this respect, its role is dual to the vector potential $A$ and, indeed, both the low-energy field theory \eqref{LL} and the clock model \eqref{Potts} emerging at large $\J$ display Kramers-Wannier dualities consistent with the exchange $A \leftrightarrow 2\pi n_g/3$ (see, for instance, Ref. \cite{Whitsitt2018}).

The role of $n_g$ is most easily understood in the Coulomb-dominated regime $\Ec \gg E_J,\J$. In this case, the JJA \eqref{hamtot} can be approximated in terms of strongly interacting Cooper pairs defining a Bose-Hubbard model \cite{Fazio2001rev}, which must be extended to account for the onsite $V(\varphi)$ potential. We stress that this mapping is different from the Ostlund construction of Eq. \eqref{hambos}; differently from the Josephson-dominated physics described in Sec. \ref{sec:class}, in the Coulomb-dominated regime $(\Ec \gg E_J,\J)$ the JJA displays two kinds of phases: the gapped Mott insulating phases appearing for $n_g$ close to integer values and the gapless superconducting phases emerging for $n_g$ close to half-integer values. 
The former are commensurate phases in which the charge fluctuations are suppressed; the expectation value of $N_i$ acquires integer values and it can be considered a winding number for the string operators $\tilde{S}$ in Eq. \eqref{string} (see Appendix \ref{app:extradata}). 
The Mott phase characterized by $\left\langle N_j \right \rangle =0$ corresponds to the disordered paramagnetic phase discussed in the previous sections, and, in general, Mott phases appear with a periodicity 1 in $n_g$. 
For $n_g = 1/2$, instead, states of the single SC islands with charge $N_j=0$ and $N_j=1$ are degenerate. 
This situation is indeed described by a Bose-Hubbard model of Cooper pairs at half-filling and
 $n_g$ becomes related to their chemical potential via \cite{Athanasiou2024}:
\begin{equation}
\mu= \Ec\left(1-2n_g\right)\,.
\end{equation}
The bandwidth of the Cooper pairs is given by $2E_J$, whereas the onsite potential $V(\varphi)$ corresponds to a term exchanging triplets of Cooper pairs with the SC background and it is strongly suppressed for large $\Ec \gg \J$.
The superfluid phase emerging in proximity of $n_g=1/2$ is captured by the Luttinger Hamiltonian \eqref{LL} when $K$ is sufficiently small such that the sine-Gordon $\cos 3 \varphi$ term is irrelevant, and $n_g$ causes a fast modulation of the field $\theta$, such that the potential $\cos 2\theta$ is strongly suppressed. This gapless phase (with $c=1$) is connected in general with the gapless incommensurate phase discussed in the previous section.

\begin{figure}
    \centering
    \includegraphics[width=\columnwidth]{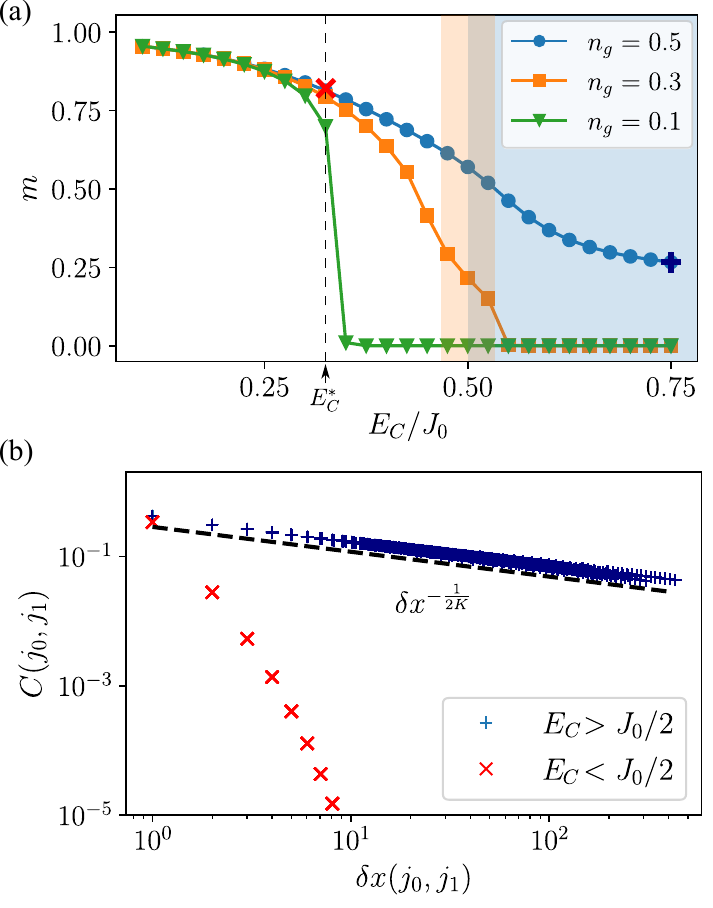}

    \caption{(a) SC order parameter as a function of $\Ec/\J$ for three values of $n_g$. As the induced charge approaches $n_g=0.5$, the critical point $\Ec^*$ increases.
    The shaded areas indicate where the correlation functions for $n_g=0.3$ and $n_g=0.5$ display a power-law decay.
    (b) connected correlation function at $n_g=0.5$ for two values of $\Ec$, corresponding to the marked data in panel (a). The power law behavior $\delta x^{-1/2K}$ with $K=1.3$ highlights the emergence of the gapless superfluid phase. Both panels have $L=100$ and $E_J=0.2\J$.
    }
    \label{fig:ng}
\end{figure}

The reduction of the Mott lobes when $n_g$ approaches 0.5 can be appreciated from Fig.~\ref{fig:ng}(a), where we show the $\mathbb{Z}_3$ order parameter $m$ as a function of the charging energy. As $n_g$ increases from 0 towards 0.5, the critical point $\Ec^*$ is also shifted towards larger values, meaning that the disordered (Mott) phase is suppressed. 
Close to $n_g=0.5$, we observe a transition between a gapped superconducting phase with $m\neq 0$ at small $\Ec$, to the gapless superfluid phase at large $\Ec$; in the latter, the order parameter decays as a power-law of the system size $L$.
Indeed, we observe markedly different behaviors between the data for $n_g=0.1$ (green triangles) and $n_g=0.5$ (blue circles). 
The former displays a standard second-order phase transition phenomenology, where the order parameter drops continuously to zero after a critical $\Ec^*(n_g>0) > \Ec^*$. 
The latter, instead, seems to saturate towards a finite value even at large $\Ec/\J$.
The data for $n_g=0.3$ (orange squares) have an intermediate behavior, where $m\to 0$ after a finite critical region, highlighted by the shaded area.

The criticality associated with the gapless superfluid phase is further confirmed by the connected correlation function in Fig.~\ref{fig:ng}(b): while for $\Ec\lesssim \J/2$ (red ``x''s) we see an exponential decay (with finite $m$), marking the ordered phase, when $\Ec\gtrsim \J/2$ (navy-blue crosses) the correlation function displays a clear power-law behavior which extends for a finite region around the charge degeneracy point.

Additionally, we observe that for non-integer values of $n_g$, the string correlators $\tilde{S}$ in Eq. \eqref{string} acquire an incommensurate winding outside the Mott/disordered phases; see Fig.~\ref{fig:winding_ng} in Appendix~\ref{app:extradata} for more details. Therefore, the gapped superconducting/ordered phases acquire an incommensurate behavior which is dual to the incommensurability discussed in the previous section.

This implies that, for $n_g \neq 0$, the transition from ordered superconducting and disordered insulating phases driven by $\Ec$ is no longer of the Potts kind, analogously to the case with generic $A$. Moreover, for generic $A$ and $n_g$, the transitions from the gapless floating phase to the gapped phases is always of the commensurate-incommensurate kind, either for the charge degrees of freedom (thus the $\theta$ field) or for the phase $\varphi$ degrees of freedom.

The duality between $n_g$ and $A$ can be made even more explicit in the clock model limit \eqref{Potts}: for large $\J \gg E_J,\Ec$ the JJA converges to the chiral clock model described in Ref. \cite{Howes1983} which, indeed, displays two kinds of gapped phases, namely the ordered and disordered phases which acquire, respectively, commensurate windings of the $\tilde{C}$ and $\tilde{S}$ correlations functions.

\section{Perturbations and disorder} \label{sec:pert_dmrg}

The analysis presented in the previous sections relies on the assumption of a perfect $\mathbb{Z}_3$ symmetry, inherited by the system from the potential \eqref{pot}. In experimental devices, however, we expect that there can be a detuning of the parameters of the triple Josephson junctions leading to perturbations of the potential $V(\varphi)$. In the following, we focus on the simplest kind of perturbations that can affect the system and we assume that all the triple junctions are affected by the same detuning resulting in the Josephson potential:
\begin{equation} \label{pert}
V(\varphi)=-\J \cos 3\varphi - \delta J_1 \cos \varphi - \delta J_2 \cos 2\varphi\,.
\end{equation}
The onset of the lower harmonics $\cos \varphi$ and $\cos 2 \varphi$ is the most fundamental perturbation due to errors in the tuning of the triple Josephson junctions and corresponds, for instance, to a displacement of its magnetic fluxes away from $\Phi_0/3$, or to a perturbation of the transparency $T_2$ away from $T_1=T_3$. The potential \eqref{pert} still preserves the inversion symmetry $\varphi \to -\varphi$, but, importantly, it breaks the $\mathbb{Z}_3$ symmetry $\varphi \to \varphi +2\pi/3$, such that its effects properly exemplify the major consequences of general perturbations independent on the position.

The primary effect of either $\delta J_1$ or $\delta J_2$ is to lift the degeneracy of the low-energy phase eigenstates $\ket{s}$. In the clock model description, their splitting corresponds to the introduction of a longitudinal magnetic term in the Hamiltonian \eqref{Potts} of the form:
\begin{equation} \label{long}
\delta H = -\delta h_\sigma \sum_j \left( \sigma_j + \sigma_j^\dag\right)\,,
\end{equation}
with $\delta h_\sigma = \delta J_1 + \delta J_2$. Additional effects are discussed in Appendix \ref{app:pert_theory}. When considering the Potts limit ($n_g=A=0$) analyzed in Sec. \ref{sec:potts}, the longitudinal term \eqref{long} is the most relevant perturbation in the proximity of the Potts critical point and it splits the degeneracy of the symmetry-broken ground states in the ordered phase. As a result, when $\delta h_\sigma \neq 0$, the system does not display any longer a quantum phase transition: the ordered and disordered phases become adiabatically connected due to the explicit symmetry breaking (see, for instance, Ref. \cite{Nyhegn2021}). This effect is analogous to the introduction of a longitudinal field in the Ising model \cite{Coldea2010,Banuls2011}.
\begin{figure}
    \centering
    \includegraphics[width=\columnwidth]{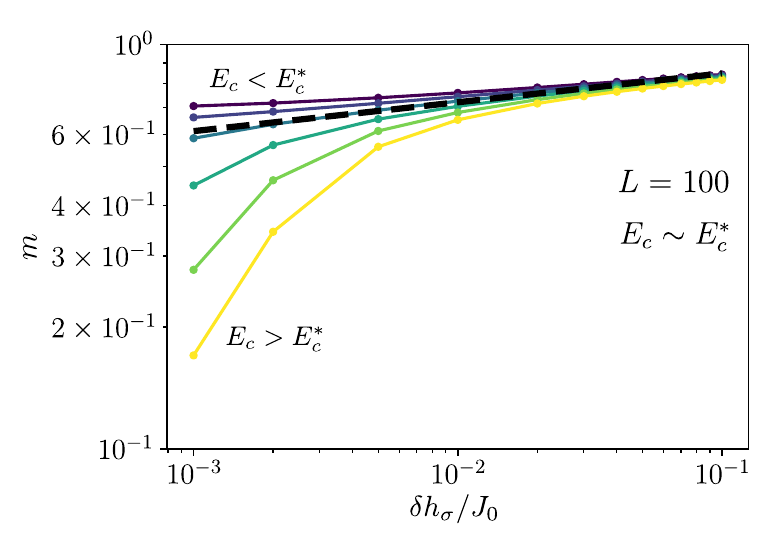}
    \caption{Order parameter as a function of the perturbation $\delta h_\sigma$ for values of $\Ec$ close to the Potts critical point $\Ec^*$. The black dashed line marks the expected scaling $m\propto \delta h_\sigma^{1/14}$.
    Here, $\delta J_1=\delta J_2 = \delta h_\sigma/2$}
    \label{fig:perturbation}
\end{figure}

The disappearance of the Potts phase transition is reflected in the onset of a finite magnetization $m$ as we perturb the critical point at $\Ec=\Ec^*$ through the introduction of $\delta J_1$ and $\delta J_2$. Figure \ref{fig:perturbation} shows the behavior of $m$ as a function of $\delta h_\sigma$ for values of $\Ec$ close to the phase transition identified in Sec. \ref{sec:potts}. The universal critical exponents of the Potts model concur in setting the scaling $m \propto \delta h_\sigma^{1/14}$ (black dashed line) for small $\delta J_1$ and $\delta J_2$.
The perturbation in Eq. \eqref{pert} affects also the value of $h_\tau$, such that a precise determination of the behavior of the magnetization as a function of $\delta J_1$ and $\delta J_2$ is non-trivial. The numerical data show that this perturbation, as expected, is weaker on the disordered side of the phase transition and gets stronger towards the ordered phase, by crossing an intermediate power-law behavior compatible with the universal exponent $1/14$. In Appendix \ref{app:pert_theory} we present a more detailed interpretation of the perturbation \eqref{pert} in terms of the perturbations of the critical clock model.

The main conclusion is that the Potts critical point is unstable under the detuning of the triple Josephson junctions, although the scaling of the superconducting order parameter $m$ as a function of $\delta J_i$ (thus, for instance, of the variation of the magnetic fluxes away from $\Phi_0/3$) can provide further evidence of the Potts universality class through the related critical exponents.

The perturbations in the potential \eqref{pert}, instead, do not affect in a significant way the appearance of the incommensurate floating phase and the phase transitions between the gapless and gapped phases.
In particular, the introduction of $\delta J_1$ and $\delta J_2$ yields simple corrections to the boundaries of the classical regimes in Fig. \ref{fig:class}, such that, for instance, the critical value of $A$ in Eq. \eqref{arccos} will be displaced by these perturbations. The boundary between the ferromagnetic and helical phases will be also linearly shifted away from $A=\pi/3$ as a function of $\delta J_1$ and $\delta J_2$. These parameters, indeed, correspond to a correction to the chemical potential $\mu$ in Eqs. \eqref{muost} and \eqref{hambos}.
Finally, we observe that perturbations of the potential $V(\varphi)$ do not play any major role in the Coulomb-dominated regime: the Mott insulator - superfluid phase transitions as a function of the induced charge $n_g$ are insensitive on the potential $V(\varphi)$ for large $\Ec$.

Another perturbation of the Hamiltonian \eqref{hamtot} that may affect the experimental platforms is the onset of repulsive interactions between the islands. As mentioned in Sec. \ref{sec:JJA}, these interactions are dictated by the inverse capacitance matrix of the system $C^{-1}$. Among them, the most relevant are given by the nearest-neighbor terms of the form $E_{cc}\left(N_j-n_g \right)\left(N_{j+1}-n_g\right)$ characterized by an energy scale $E_{cc}<\Ec/2$ (in particular, for hybrid systems, $E_{cc}< \Ec/4$ in the realistic estimates presented in Ref. \cite{Materise2023}). 
These mutual electrostatic interactions do not lead to qualitative differences in the physics of the Josephson junction arrays for systems that are dominated by the Josephson potentials. In particular, nearest-neighbor interactions can be neglected in the transmon-like regime discussed in Sec. \ref{sec:TJJ} as the related two-island phase slips ($\tau_j \tau_{j+1}$) display amplitudes that are exponentially suppressed approximately by the factor $\exp \left[{-\sqrt{\frac{256J_0\cos(3A/2)}{9(E_c + 2E_{cc})}}}\right]$, to be compared to $h_\tau$ in Eq. \eqref{slip} (see Appendix \ref{app:WKB}).
We observe, however, that the mutual repulsive interactions lower the value of the Luttinger parameter $K$, thus lowering the critical value $\Ec^*$ and shrinking the extension of the incommensurate phase appearing around $A=\pi/3$ for low $\Ec$. 
Nearest-neighbor interactions, instead, may play a more decisive role in the Coulomb-dominated regimes. In particular they favor the onset of staggered charge density wave for $n_g\sim 1/2$ in the Bose-Hubbard model of Cooper pairs emerging for $E_c \gg E_J,J_0$ \cite{Glazman1997}. Analogously to the Mott insulating states, these staggered states suppress the charge fluctuations and constitute insulating phases that compete with 
the gapless superfluid phase described in Sec. \ref{sec:ng}. 

Finally, to provide rigorous modeling of the experimental platforms for the observation of these phases of matter, one needs to account also for the unavoidable disorder that affects the parameters of the JJAs, despite the accuracy of modern fabrication techniques.

Disorder may have a major influence when approaching the Potts critical point in the limit $n_g=A=0$. In this scenario, we must distinguish two kinds of perturbations: perturbations that leave the system invariant under the $\mathbb{Z}_3$ symmetry, such as disorder in the $E_J,\J,\Ec,A$ and $n_g$ parameters, may lead to relatively weak deformations of the low-energy scaling limit of the model since they are coupled to operators with larger scaling dimensions based on the scaling analysis of disordered systems \cite{Giamarchi1987}. Specifically, quenched disorder in the parameters $M$ and $\J$ in Eq. \eqref{LL} can lead to different scaling exponents similarly to a random transverse field for the quantum Ising model \cite{Fisher1992}. On the other hand, perturbations that break the $\mathbb{Z}_3$ symmetry, instead, are expected to have an effect similar to the detuning of the single island potential $\delta J_i$ discussed above. In this respect, disorder is expected to introduce an additional length scale in the model, which diverges in the clean systems. The Potts critical behavior can be observed only if this length scale is sufficiently large to be comparable with the system size. A complete understanding of the role of disorder at the Potts transition, however, requires a deep analysis in terms of renormalization group and CFTs and goes beyond the scope of this work.

The role of disorder becomes even more complex when considering the commensurate-incommensurate phase transitions and the floating phase. For instance, in the regime of low $\Ec$ at $A\sim \pi/3$, we expect that charge disorder is strongly suppressed, and the incommensurability introduced by $A$ additionally suppresses the disorder in the onsite Josephson potential. In this case, it may even be possible that a further disorder in $\J$ or in the parameter $A$ could extend the floating phase. Concerning the Coulomb-dominated regime at large $\Ec$, instead, the role of charge disorder becomes important; its effect can be understood based on the mapping to the Bose-Hubbard model and can lead to the formation of additional Bose glass phases \cite{Giamarchi1988} (see also the numerical results in Ref. \cite{Gerster2016}).

As recent experiments in long JJAs~\cite{manucharyan2019} estimated a disorder strength in the Josephson couplings below 10\%, however, we are confident that a state-of-the-art fabrication of the device allows for the observation of all the main features of the phases of matter discussed in the previous sections.

\section{Conclusions}\label{sec:conclusion}
In this paper, we proposed a platform based on tunable one-dimensional Josephson junction arrays to perform analog quantum simulations of one-dimensional quantum clock models. 
Its building blocks are triple Josephson junctions tuned in a regime where their energy-phase relation is characterized by three degenerate minima related by a $\mathbb{Z}_3$ symmetry.

These superconducting systems provide a solid-state architecture for quantum simulations that combines two key features: (i) their dynamics is determined by two controllable incommensurability parameters, which correspond to the induced charges in the superconducting islands and the magnetic fluxes in their geometry; (ii) they realize a global discrete $\mathbb{Z}_3$ symmetry which, differently from Rydberg atom experiments \cite{bernien2017,Keesling2019} and triangular lattice antiferromagnets \cite{Chen2024}, is not related to spacial symmetries as translational or rotational invariance. 
The $\mathbb{Z}_3$ symmetric and symmetry-broken states of these Josephson junction chains are indeed associated with different configurations of their superconducting phases. This symmetry can be engineered already at the level of a single Josephson junction element by taking advantage of the gate-tunability of the energy-phase relation recently demonstrated in hybrid parallel junctions and loops\cite{Banszerus2024,banszerus2024b}. 

In the many-body scenario we analyzed, we showed through detailed DMRG analysis and effective field-theoretical approaches that hybrid superconducting circuits provide a versatile platform to realize the scaling limit of interacting bosonic quantum field theories \cite{Roy2019,Saleur2021} also in the presence of discrete symmetries.
The Josephson junction chain we proposed has a very rich phase diagram that includes nontrivial critical states, and, in particular, both the conformal Potts phase transitions and gapless floating phases which extend even to the classical limit of this superconducting array. For these reasons, it provides a solid-state alternative to Rydberg atom setups to implement controllable quantum simulations of commensurate/incommensurate phase transitions.

In a broader context, hybrid JJAs are compelling elements for pushing superconducting quantum circuits beyond the qubit realm. 
We observe, in particular, that the triple Josephson junction we described can be operated as a qutrit, if suitably coupled with a microwave resonator as for standard transmons. In this scenario, the computational states are encoded in the minima of the Josephson potential as in parity~\cite{Schrade_PRXQ2022} and modern flux qubits~\cite{Flensberg2024}. 
Single qutrit-gates could be straightly generalized from their qubit counterparts~\cite{Flensberg2024,Krantz2019}, while entangling gates and readout methods would require a more careful investigation. In this respect, the superconducting arrays that we discussed constitute a starting point for the realization of tunable solid-state qutrit architectures.

\begin{acknowledgements}
We warmly thank E. Cobanera, L. Banszerus, G. Ortiz, M. Rizzi, N. Tausendpfund, and S. Vaitiekenas for useful discussions.

MW has received funding from the European Union’s Horizon Europe research and innovation programme under grant agreement No 101080086 NeQST.
Views and opinions expressed are however those of the author(s) only and do not necessarily reflect those of the European Union or the European Commission. Neither the European Union nor the granting authority can be held responsible for them.
This work was supported by Q@TN, the joint lab between the University of Trento, FBK—Fondazione Bruno Kessler, INFN—National Institute for Nuclear Physics, and CNR—National Research Council. M.B. and L.M. have been supported by the Villum Foundation (Research Grant No. 25310). L.M. has been supported by the research grant ``PARD 2023'', and ``Progetto di Eccellenza 23-27'' funded by the Department of Physics and Astronomy G. Galilei, University of Padua.

\end{acknowledgements}

\section*{Data Availability}
The data presented in this article is available from~\cite{Wauters_Zenodo2025}.

\appendix

\section{Semiclassical estimate of the phase slips in the triple junction} \label{app:WKB}

The estimate of the amplitude of the $2\pi/3$ phase slip processes that determine the dynamics described in Eq. \eqref{slip} can be obtained by a standard WKB approximation that follows the analogous procedures for transmon devices \cite{Koch2007}. From the Hamiltonian \eqref{ham2} of a triple-junction building block in the transmon regime $(\J \gg \Ec)$ we derive a semiclassic momentum expressed as:
\begin{equation}
-i \partial_\varphi = \sqrt{-\frac{\J}{\Ec}\left(1-\cos 3 \varphi\right)} +n_g\,.
\end{equation}
Consequently, the exponential suppression of the tunneling amplitude is given by:
\begin{equation} \label{WKB}
\ee^{-\int_0^{\frac{2\pi}{3}} d\varphi \sqrt{\frac{\J}{\Ec} \left(1-\cos 3 \varphi\right)} + i\frac{2\pi}{3}n_g} = \ee^{- \sqrt{\frac{32\J}{9\Ec}} + i\frac{2\pi}{3}n_g}\,.
\end{equation}
Concerning the prefactor in Eq. \eqref{slip}, we consider the variable redefinition $\varphi' = 3\varphi$, such that:
\begin{equation}
H'=9\Ec\left(-i\partial_{\varphi'} - \frac{n_g}{3}\right)^2 + \J\left(1-\cos \varphi'\right)\,.
\end{equation} 
Therefore, to map the triple junction phase slips into the standard phase slips in a transmon \cite{Koch2007}, we must simply rescale $\Ec \to \frac{9}{4}\Ec$. Consequently the prefactor in Eq. \eqref{slip} corresponds to:
\begin{equation} \label{alpha}
\alpha \J^{3/4}\Ec^{1/4} \equiv 8\J^{3/4}\Ec^{1/4}  2^{5/4} \sqrt{\frac{3}{\pi}} \approx 18.6 \J^{3/4}\Ec^{1/4}\,.
\end{equation}
In our mapping to the quantum clock model, we neglected the simultaneous phase slips of neighboring islands.
Analogously to the physics of Josephson junction chains hosting $\mathbb{Z}_N$ parafermions \cite{Milsted2014,Wouters2022}, these two-island phase slips would amount to operators of the form $\tau_j \tau_{j+1}$ in the clock model limit. They arise, in particular, when neighboring islands are interacting through an electrostatic repulsion of the form $E_{cc}\left(N_j-n_g \right)\left(N_{j+1}-n_g\right)$. 

The amplitude of these two-island phase slips can be estimated by adopting the approach in Refs. \cite{Hassler2012,Nielsen2023}. In particular, it can be easily evaluated in the limit $E_J > J_0 \gg \Ec > E_{cc}$. 
In this scenario, we can consider two neighboring islands as a single superconducting element sharing the same phase. Such element acquires an effective charging energy $\tilde{E}_c = \frac{2\Ec + E_{cc}}{4}< \Ec$ and is connected to the superconducting background through an effective SQUID for triplets of Cooper pairs with potential $-\tilde{J}_0\cos 3 \varphi$, with $\tilde{J}_0 = 2J_0\cos(3A/2)$\,.

By repeating the previous calculations, we derive that the amplitude of the two-island phase slips scales as $\tilde{J}_0^{3/4}\tilde{E}_c^{1/3}\ee^{-\sqrt{\frac{32\tilde{J}_0}{9\tilde{E}_c}}}$. For small $A$, this amplitude is strongly suppressed with respect to $h_\tau$ in Eq. \eqref{WKB}, thus bearing only minor corrections to the Potts phase transition.

We emphasize that our numerical MPS simulations do not rely on any of these semiclassical approximations, and capture the ground states of the system independently on the ratio $\J/\Ec$.

\section{Kink mass in the effective field theory description}\label{sec:kink}

The effective field theory approximation \eqref{hambos} includes static kink solutions for the dual field $\theta\left(x,t\right)$. A rough estimate of their mass can be obtained by considering the limiting case $\Ec \gg J_0, E_J$ and following the standard calculations (see for example Ref.\cite{mussardobook}).

In the Coulomb-dominated regime, the energy of a generic configuration $\theta\left(x,t\right)$ can be written as
\begin{multline} \label{etheta}
    E\left[\theta\right] = \dfrac{1}{2\pi K}\int dx \left[\dfrac{\left(\partial_t\theta\right)^2}{v}+v\left(\partial_x\theta-\dfrac{\pi n_g}{a}\right)^2\right]+\\
    -M\int dx \cos{\left(2\theta\right)}.
\end{multline}
The kink corresponds to a static solution $\theta(x,t)=\bar{\theta}(x)$ of the equation of motion,
\begin{equation} \label{kinkeom}
    \dfrac{\partial^2\theta}{\partial t^2} - v^2\dfrac{\partial^2\theta}{\partial x^2} = -2MvK\pi\sin(2\theta).
\end{equation}
which interpolates between two minima of the potential $V(\theta) = - M \cos(2\theta)$.
The induced charge $n_g$ gives rise just to a linear term in the space derivative in Eq. \eqref{etheta} and therefore does not appear in the equation of motion.
The two static solutions of Eq. \eqref{kinkeom},
\begin{equation}
    \bar{\theta}_{\pm}(x)=2\arctan{\left[\ee^{\pm 2\sqrt{\pi M K/v}x}\right]}\,,
\end{equation}
thus describe the kink and anti-kink excitations with finite energy
\begin{equation}
    E\left[\bar{\theta}_{\pm}(x)\right]= 4 \sqrt{\dfrac{Mv}{\pi K}}\mp\dfrac{\pi v}{a K} n_g.
\end{equation}
Replacing the perturbative expression \eqref{eq:luttinger_p} for the Luttinger liquid parameters yields the mass of the kink $M_{\rm kink}$ adopted in the main text, from which we derive the parameter $M$ in Eq. \eqref{Mparameter}.

\section{Further numerical results}\label{app:extradata}
\subsection{Phase diagram for $\J<E_J/2$}\label{app:extradata_phasediagr}
\begin{figure}
    \centering
    \includegraphics[width=\columnwidth]{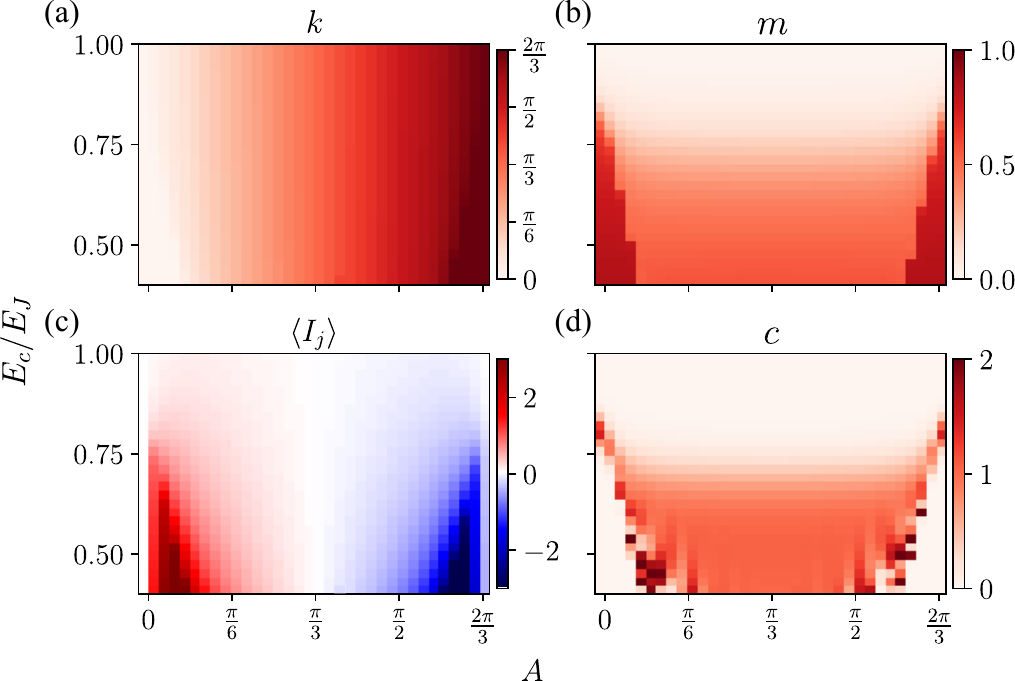}
    \caption{Phase diagram for $E_J=2.5\J$, $L=100$. (a) winding $k$ of the correlation function $C(j_0,j_1)$; (b) $\mathbb{Z}_3$ order parameter $m$; (c) current; (d) central charge $c$. The gapless phases is much more extended compared to Fig.~\ref{fig:phase_diag}}
    \label{fig:phase_diag2}
\end{figure}

We report in Fig.~\ref{fig:phase_diag2} the phase diagram in the $\Ec - A $ plane when the nearest-neighbour Josephson coupling is the dominant energy scale $E_J=2.5\J>\Ec$.
We observe the same overall structure of Fig.~\ref{fig:phase_diag}, with the difference that now the gapless phase, marked by the central charge $c\simeq 1$, extends over a much larger region.

\subsection{Winding of $\tilde{S}$ at finite $n_g$}
\begin{figure}
    \centering
    \includegraphics[width=\columnwidth]{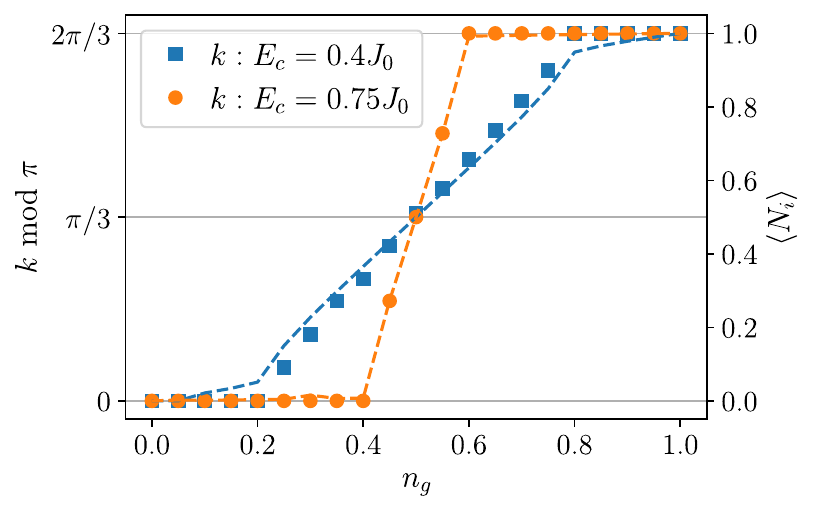}
    \caption{Momentum $k$ associated to the phase modulation of $\tilde{S}$ as a function of the induced charge. The dashed lines represent the average charge on the SC islands (right y-axis). Notice the similarity with Fig.~\ref{fig:flux}(a), which highlights the duality between $n_g$ and $A$. Simulation parameters: $L=100$, $E_J=0.2\J$.}
    \label{fig:winding_ng}
\end{figure}

As discussed in Sec.~\ref{sec:ng}, the induced charge is dual to the Peirels phase $A$ acquired by Cooper pairs hopping between neighboring SC islands.
To appreciate this, we report in Fig.~\ref{fig:winding_ng} the behavior of the string correlations $\tilde{S}(j_0,j_1)$ for finite $n_g$ and $\Ec<\J/2$, so that the system is still in the ordered phase.
The amplitude of $\tilde{S}$ decays exponentially with the distance, modulo boundary effects, as clearly shown in panel (a).
However, it acquires a nontrivial incommensurate phase modulation $\ee^{-ik|j_1-j_0|}$.
We summarize its behavior in Fig.~\ref{fig:winding_ng}.
Analogously to Fig.~\ref{fig:flux}(a) for the phase $A$, we observe three regimes as $n_g$ is tuned from 0 to 1.
For $n_g<n_g^*$, the string correlation displays a staggered phase modulation associated with $k=\pi$.
To facilitate the visualization of the modulation of $k$ on top of this staggering, we remove it by showing $k \mod \pi.$
This region corresponds to the Mott insulator and its extension increases for larger charging energies.
For $n_g>1-n_g^*$, the phase winding $k=2\pi/3$ is commensurate with the lattice spacing and this ``helical'' Mott insulator is dual to the SC helical phase in Fig.~\ref{fig:phase_diag}.
Finally, between $n_g^*$ and $1-n_g^*$, the phase winding $k$ is incommensurate and connects the two plateaus. In this region both the gapped SC($\Ec < \J/2)$ and gapless superfluid ($\Ec>\J)$ appear.
The overall behavior of the phase winding is described by 
\begin{equation}\label{eq:winding_ng}
 k\simeq \frac{2\pi}{3}\langle N_i\rangle \ ,   
\end{equation}
where $\langle N_i \rangle $ is the average charge on the SC islands, as shown by the excellent correspondence between $k$ (markers) and $\langle N_i \rangle $ (dashed lines) in Fig.~\ref{fig:winding_ng}.

\section{Perturbations of the single-island potential} \label{app:pert_theory}

In Sec. \ref{sec:pert_dmrg} we discussed the main effect of the perturbations $\delta J_i$ in the potential \eqref{pert} in terms of a splitting in energy of the semiclassical potential minima that correspond to the longitudinal field in Eq. \eqref{long}.

The corrections $\delta J_i$, however, yield additional effects that can be captured in terms of modifications of the amplitude of the phase slips presented in Appendix \ref{app:WKB}. In particular, the lower harmonics in $V(\varphi)$ break the equivalence of the transitions between the three minima potential minima such that the transitions between $\varphi=0$ and $\varphi= \pm 2\pi/3$ acquire a different amplitude with respect to the transitions between $2\pi/3$ and $-2\pi/3$. 

To account for this effect, we can re-estimate the amplitudes in Eq. \eqref{WKB} separately for the three transitions. We focus, in particular, on the transmon regime (large $\J$) such that our analysis can be performed by considering the emerging clock degrees of freedom. 

The corrected amplitudes of the phase slips as a function of $\delta J_1$ and $\delta J_2$ determine on one side a general change of the transverse field by a quantity $\delta h_\tau$ and, on the other, the addition of a perturbation that can be expressed as a function of products of $\sigma$ and $\tau$ onsite. In particular, the onsite transverse field of the clock model must be modified by the introduction of the following terms:
\begin{equation}
 \delta H' = -\delta h_\tau \left(\ee^{i n_g \frac{2\pi}{3}} \tau +  \ee^{-i n_g \frac {2\pi}{3}} \tau^\dag\right) - \delta h_{\sigma\tau} f(\tau,\sigma)\,,
\end{equation}
with
\begin{equation}
f(\tau,\sigma)\equiv \ee^{-i\left(n_g+\frac{1}{2}\right) \frac{2\pi}{3}} \left(\tau\sigma+\sigma^\dag \tau \right) + {\rm H.c.}  \,,
\end{equation}
where this last term is obtained by observing that:
\begin{equation} \label{perttau}
e^{-i\frac{\pi}{3}}\left(\tau\sigma+\sigma^\dag \tau \right) = \begin{pmatrix} 0 & 1 & 0 \\ 0 & 0 & -2 \\ 1 & 0 & 0\end{pmatrix}\,.
\end{equation}
The last matrix makes it explicit that the transition amplitude between $\varphi=0$ and $\varphi= \pm 2\pi/3$ is different from the transition amplitude between $2\pi/3$ and $-2 \pi/3$. 

These amplitudes can be approximated by introducing two related coefficients, $\alpha_{0,2\pi/3}$ and $\alpha_{2\pi/3,-2\pi/3}$, which correspond to the perturbed version of the parameter $\alpha$ in Eqs. \eqref{slip} and \eqref{alpha}. Their differences with $\alpha$ are linear in $\delta J_i / \J$. By numerically estimating these coefficients, one can derive the parameters $\delta h_\tau$ and $\delta h_{\sigma \tau}$ through the equations:
\begin{align}
&\approx \frac{h_\tau}{\alpha} \left(\frac{2}{3}\alpha_{0,2\pi/3} + \frac{1}{3}\alpha_{2\pi/3,-2\pi/3} -\alpha \right) ,\\
& \delta h_{\sigma \tau} \approx \frac{h_\tau}{3\alpha} \left(\alpha_{0,2\pi/3} - \alpha_{2\pi/3,-2\pi/3}\right)\,.
\end{align}
Considering the unperturbed system at the Potts phase transition, the perturbations $\delta h_\tau$ and $\delta h_{\sigma \tau}$ are relevant but subleading with respect to the longitudinal field term with coefficient $\delta h_\sigma$. In general, we expect a scaling of the order parameter away from the critical point dictated by the universal behavior:
\begin{equation}
m = \sum_a A_a \delta h_a^{\frac{D_\sigma}{2-D_a}}\,,
\end{equation}
where $a$ runs over the relevant primary perturbations determined by the Potts conformal field theory, $D_a$ are the scaling dimensions of the corresponding primary fields and $A_a$ are non-universal constants. When considering the mapping of the clock model into the Potts conformal field theory (see Ref. \cite{Mong2014} for advanced analysis), the perturbation in $\delta h_\sigma$ matches the primary field with dimension $D_\sigma = \frac{2}{15}$ and the displacement $\delta h_\tau$ corresponds to the primary field with dimension $D_\tau=4/5$. The perturbation $\delta h_{\sigma \tau}$, instead, is not directly related to a primary operator. Its dominant term should however be characterized by the same scaling of the longitudinal field.

In general, for small values of $\delta J_i$, the term $\delta h_\sigma$ is therefore the most relevant and determines the critical exponent $1/14$ considered in Fig. \ref{fig:perturbation}. The deviations from this scaling can be caused by the influence of the subleading contributions and the fact that our numerics does not exactly sit at the critical point even in the unperturbed limit.

%

\end{document}